\documentclass[iop]{emulateapj}
\usepackage{apjfonts}
\usepackage{amsmath,amstext}
\usepackage{booktabs,pbox}
\usepackage{epsfig,graphicx}
\usepackage[breaklinks,colorlinks,citecolor=blue,linkcolor=magenta]{hyperref}
\usepackage{natbib}
\usepackage{color}

\slugcomment{Draft}

\shorttitle{A Characteristic Mass Scale in the MZR}
\shortauthors{Blanc et al. }

\begin{document}

\title{A Characteristic Mass Scale in the 
Mass-Metallicity Relation of Galaxies}

\author{Guillermo A. Blanc\altaffilmark{1,2,3}, Yu Lu\altaffilmark{1},
  Andrew Benson\altaffilmark{1}, Antonios Katsianis\altaffilmark{2}, Marcelo Barraza\altaffilmark{2,4}}

\altaffiltext{1}{Observatories of the Carnegie
  Institution for Science, 813 Santa Barbara St., Pasadena, CA, USA}
\altaffiltext{2}{Departamento de Astronom\'ia, Universidad de Chile,
  Camino del Observatorio 1515, Las Condes, Santiago, Chile}
\altaffiltext{3}{Centro de Astrof\'isica y Tecnolog\'ias Afines (CATA),
  Camino del Observatorio 1515, Las Condes, Santiago, Chile}
\altaffiltext{4}{Max Planck Institut f\"ur Astronomie, K\"onigstuhl 17, D-69117 Heidelberg, Germany}

\begin{abstract}

We study the shape of the gas-phase mass-metallicity relation (MZR) of a combined sample of present-day dwarf and high-mass star-forming galaxies using {\tt IZI}, a Bayesian formalism for measuring chemical abundances presented in \cite{blanc15}. We observe a characteristic stellar mass scale at $M_* \simeq 10^{9.5}$~M$_{\odot}$, above which the ISM undergoes a sharp increase in its level of chemical enrichment. In the $10^{6}-10^{9.5}$~M$_{\odot}$ range the MZR follows a shallow power-law ($Z\propto M^{\alpha}_*$) with slope $\alpha=0.14\pm0.08$. At approaching $M_* \simeq 10^{9.5}$~M$_{\odot}$ the MZR steepens significantly, showing a slope of $\alpha=0.37\pm0.08$ in the $10^{9.5}-10^{10.5}$~M$_{\odot}$ range, and a flattening towards a constant metallicity at higher stellar masses. This behavior is qualitatively different from results in the literature that show a single power-law MZR towards the low mass end. We thoroughly explore systematic uncertainties in our measurement, and show that the shape of the MZR is not induced by sample selection, aperture effects, a changing N/O abundance, the adopted methodology to construct the MZR, secondary dependencies on star formation activity, nor diffuse ionized gas (DIG) contamination, but rather on differences in the method used to measure abundances. High resolution hydrodynamical simulations of galaxies can qualitatively reproduce our result, and suggest a transition in the ability of galaxies to retain their metals for stellar masses above this threshold. The MZR characteristic mass scale also coincides with a transition in the scale height and clumpiness of cold gas disks, and a typical gas fraction below which the efficiency of star formation feedback for driving outflows is expected to decrease sharply. 

\end{abstract}


\section{Introduction}\label{sec:introduction}

The chemical enrichment of the inter-stellar medium (ISM) of star forming galaxies is modulated by gas accretion from the inter-galactic medium (IGM), the injection and mixing of metals produced by star formation, their removal from the ISM by locking them into long lived stars and stellar remnants, their ejection out of the cold disk by means of galactic fountains and outflows, the mixing of enriched and low metallicity gas in the circum-galactic medium (CGM), and the removal or re-accretion of CGM gas out of the halo or back into galaxies \citep{lynden-bell75, larson76, lacey85, edmunds90, dalcanton07, oppenheimer10, lilly13, ma16}. All these processes play an important role in shaping how galaxy evolution proceeds. Therefore, quantifying the metal content of the ISM of galaxies provides a way to study these processes. 

In particular the stellar mass ($M_*$) vs. gas-phase metallicity ($Z$) relation \citep[hereafter MZR,][]{lequeux79, tremonti04} depends on how gas accretion, outflows, and star formation proceed as galaxies grow. It therefore carries important information about these processes. Many measurements of the gas-phase MZR have been published using samples of massive star forming galaxies at different redshifts \citep[$0<z<3.5$, e.g.][]{tremonti04, kewley08, erb06, maiolino08, zahid11, zahid14b, henry13, maier14, steidel14, sanders15}, while a few studies, mostly at $z\simeq 0$, have tried to extend the MZR to the low mass ``dwarf" galaxy regime \citep[e.g.][]{lequeux79, lee06, vaduvescu07, zahid12, andrews13}. The closely related luminosity-metallicity (LZR) relation of local dwarf galaxies has also been studied by several authors \citep[e.g.][]{lequeux79, richer95, melbourne02, salzer05, rosenberg06, sweet14}

A general trend of decreasing metallicity towards lower stellar masses is seen in all studies. But the exact form of the MZR is currently not well established because of to the presence of strong systematic uncertainties affecting metallicity diagnostics. As shown by \cite{kewley08} using a sample of galaxies from the Sloan Digital Sky Survey \citep[SDSS][]{york00}, the shape and normalization of the gas-phase MZR depends strongly on the method used to measure the oxygen abundance ($\log{\rm{O/H}}$, typically assumed to trace the ISM metallicity). At fixed stellar mass \cite{kewley08} show differences of up to 0.7~dex in the abundances derived from different strong-emission-line (SEL) methods. These systematic offsets are not constant as a function of stellar mass and translate also into significant differences in the shape of the MZR depending on the adopted diagnostics.

Possible origins for these discrepancies are extensively discussed in the literature \citep{stasinska02, stasinska04, kewley08, lopez-sanchez12, blanc15, bresolin16}, and we refer the reader to these articles for a more in depth discussion. The fact is that these systematic uncertainties directly affect our ability to extract meaningful information from the MZR regarding the physics of gas accretion, outflows, and star formation. For example, in \cite{lu15a} we use observations of the present day stellar and gas-phase MZR as well as the redshift evolution of the gas MZR to constraint the stellar mass dependence of the mass loading factor ($\eta=\dot{M}_{outflow}/SFR$), which relates the ISM outflow rate to the SFR of galaxies. In that work we find that the mass loading factor increases towards lower stellar masses, in agreement with previous studies \citep[e.g.][]{spitoni10, peeples11, muratov15, christensen16, christensen18}, and the assumptions typically adopted in cosmological simulations that reproduce the galaxy stellar mass and star formation rate functions \citep[e.g.][]{springel03, dave11b, schaye15, katsianis17}. But the range of values consistent with observations is large, mostly due to systematic discrepancies between the diagnostics used for different measurements of the MZR \citep[e.g.][]{tremonti04, lee06, maiolino08, andrews13, zahid14a}. Improving the accuracy of methods to measure chemical abundances in the ISM can directly translate into better quantitative constraints on the baryon cycle of star forming galaxies.

Of particular interest is the behavior of the slope of the MZR across different regimes in stellar mass. Changes in the MZR slope can indicate changes in how the physical processes that drive galaxy evolution operate and interact as galaxies grow in mass. Most studies of the MZR find a flattening in the relation at the high mass end of the galaxy population ($M_*>10^{10.5}$~M$_{\odot}$), independently from the diagnostics used. On the other hand, the slope of the MZR in the intermediate and low mass regimes is poorly constrained and subject to strong systematics. Conflicting results exist in the literature, with some studies finding a continuous relation well modeled by a single power-law extending to low stellar masses ($\sim 10^{6-7}$~M$_{\odot}$), and others finding a mass dependent slope which is different for normal and low mass/luminosity dwarf galaxies (see references and discussion in Section \ref{sec:discussion-literature}).

In \cite{blanc15} we presented {\tt IZI}\footnote{{\tt IZI} stands for ``{\bf I}nferring metallicities ({$\bf{Z}$}) and {\bf I}onization parameters'', and the code is publicly available at http://users.obs.carnegiescience.edu/gblancm/izi}, a new method for measuring the metallicities ($Z$) and ionization parameters ($q$) of ionized nebula. The method uses Bayesian inference to quantify the posterior probability density function (PDF) of $Z$ and $q$ by comparing strong emission line flux measurements to theoretical photo-ionization models. In particular, in \cite{blanc15} we show that {\tt IZI} can provide reliable gas-phase metallicities that are in agreement (to $\sim0.1$~dex) with metal recombination line (RL) based abundance measurements in bright local HII regions. The latter being considered a ``gold standard'' for metal abundances (see discussion in \citealt{blanc15}, although c.f. recent work by \citealt{bresolin16} and the discussion in Section \ref{sec:discussion-RL}). 

In this work we measure the present day gas-phase MZR of star forming galaxies using {\tt IZI}. This new measurement is based on a literature compilation of dwarf and massive galaxy samples, spanning 5 decades in stellar mass across the $10^6$~M$_{\odot}<M_*<10^{11}$~M$_{\odot}$ range. {\tt IZI} reveals the presence of a break in the MZR at around $M_*\simeq 10^{9.5}$~M$_{\odot}$, above which the ISM of galaxies suffers a sharp increase in its level of chemical enrichment. In the $10^{6}-10^{9.5}$~M$_{\odot}$ range the MZR follows a shallow power-law of the form $Z\propto M^{\alpha}_*$, with slope $\alpha=0.14\pm0.08$. At approaching $M_* \simeq 10^{9.8}$~M$_{\odot}$ the MZR steepens significantly, showing a slope of $\alpha=0.37\pm0.08$ in the $10^{9.5}-10^{10.5}$~M$_{\odot}$ range, and a flattening towards a constant metallicity at higher stellar masses.

In Section \ref{sec:methods} we describe the samples of galaxies we use, and our measurements of the gas-phase metallicity PDFs of individual systems using {\tt IZI}. We present the resulting MZR in Section \ref{sec:results} and discuss the presence of the observed transition mass scale at $M_*\simeq 10^{9.5}$~M$_{\odot}$. In Section \ref{sec:systematics}, we explore the impact of possible sources of systematic error in the measurement, including sample selection, spectroscopic aperture effects, a changing N/O abundance in low mass galaxies, the statistical methods adopted to characterize the metallicity PDFs and build the MZR, a secondary dependence of the MZR on star formation activity, and the effects of HII region ensemble averaging and DIG contamination. We conclude that the measured MZR is robust against all the explored systematics. 

We find that the differences with previous measurements arise directly from systematic differences between the adopted abundance diagnostics. We conclude that the good ($<0.1$~dex) agreement between {\tt IZI} abundances and RL abundance measurements in local HII regions, supports the idea that the reported transition mass scale in the MZR is real. We discuss these results in Section \ref{sec:discussion} where we compare our findings with previous work in the literature and the results of state of the art cosmological and zoom-in hydrodynamical galaxy simulations. 

In Section \ref{sec:discussion} we also discuss possibilities for the physical origin of the observed transition. The MZR characteristic mass scale coincides with a threshold mass above which galaxies in zoom-in numerical simulations become very effective at retaining the metals they produce during their lives. This mass also corresponds to a transition in the scale height and clumpiness of galactic cold gas disks \citep{dalcanton04}, and a typical galaxy gas fraction below which the efficiency of star formation feedback for driving outflows is expected to decrease sharply \citep{hayward17}. This points to a scenario in which the observed break in the MZR is related to a transition in the structure of the cold ISM of star forming galaxies, and the way in which star formation feedback couples to the gas in order to eject metals outside the halo. This new measurement of the MZR has important implications for the way in which baryons are cycled between the ISM, CGM, and IGM, in and around galaxies, and how outflows are driven from these systems. Finally, we provide a summary and our final conclusions in Section \ref{sec:summary}.

\section{Methods}\label{sec:methods}

\subsection{Data and Sample Selection}
\label{sec:methods-samples}

\begin{figure}[t!]
\begin{center}
\epsscale{1.2}
\plotone{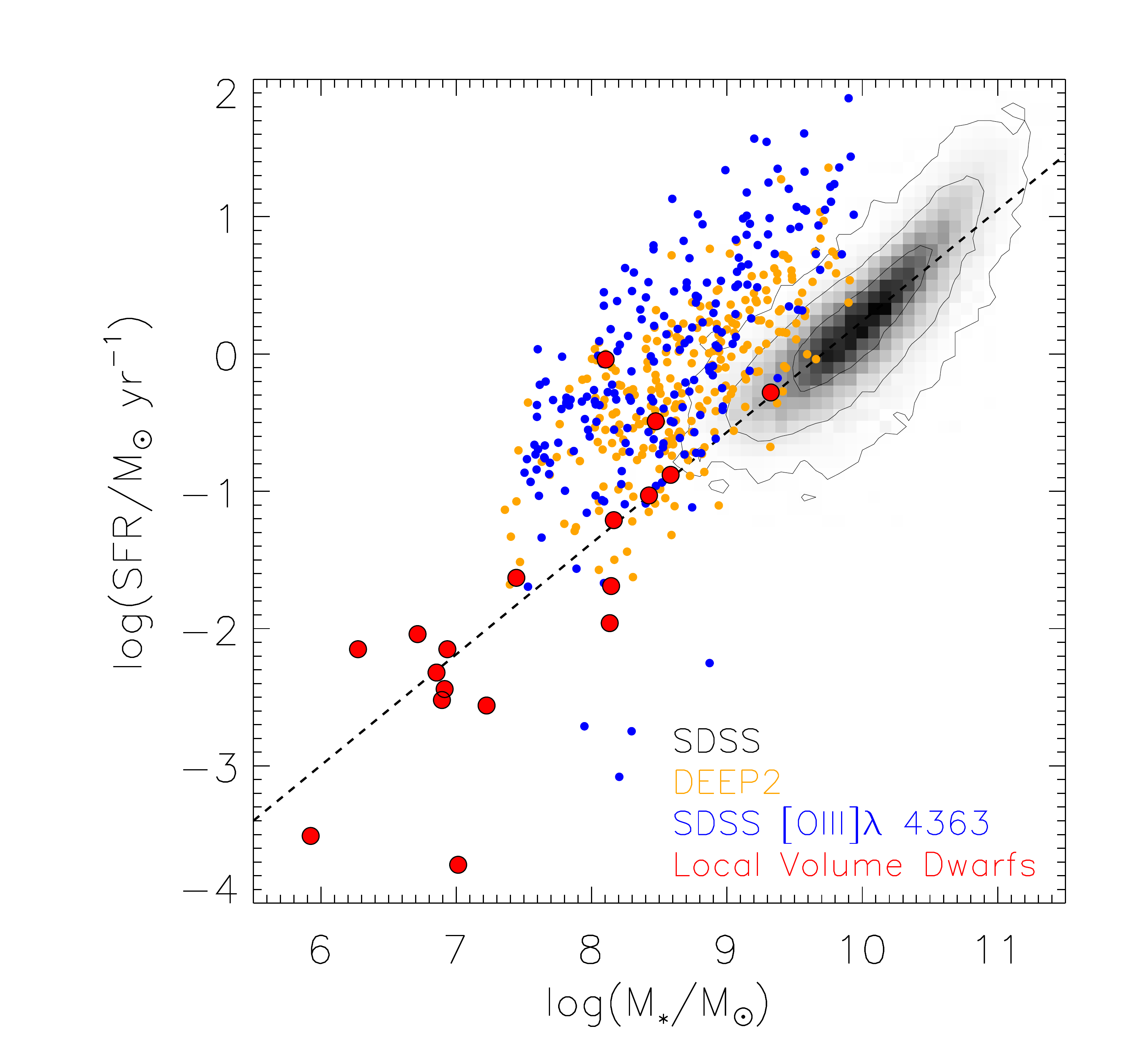}
\caption{Distribution of adopted galaxy samples on the stellar mass versus SFR plane. Gray density contours and the dashed black line show the distribution of high mass SDSS galaxies and the best-fit linear relation to the distribution. Individual galaxies in the SDSS low mass \citep{izotov06}, DEEP2 \citep{zahid12}, and Local Volume Dwarfs \citep{lee06} samples are shown by the blue, orange, and red points respectively. All stellar masses and SFRs have been scaled to a Kroupa IMF \citep{kroupa01}.}
\label{fig:m-sfr}
\end{center}
\end{figure}

To measure the gas-phase MZR over a large dynamic range in stellar mass, we analyze a representative sample of high mass star forming galaxies and three differently selected samples of low mass systems. The high mass sample is selected from the SDSS Data Release 7 MPA/JHU catalog\footnote{http://www.mpa-garching.mpg.de/SDSS/index.html} \citep{brinchmann04, abazajian09}. The three dwarf galaxy samples used in this work are taken from \cite{lee06}, \cite{izotov06}, and \cite{zahid12}. Figure \ref{fig:m-sfr} shows the distribution of the four samples in the $M^*$-SFR plane after scaling all stellar masses and SFRs to a common Kroupa IMF \citep{kroupa01}.

\subsubsection{SDSS High Mass Sample}
\label{sec:methods-samples-sdss}

The SDSS MPA/JHU catalog includes optical emission line flux measurements and associated errors from fiber spectroscopy of the SDSS DR7 ``main galaxy'' sample \citep{strauss02}, as well as stellar mass and SFR estimates derived using the methods of \cite{kauffmann03} and \cite{salim07}. In this work we use the updated MPA/JHU stellar masses\footnote{http://home.strw.leidenuniv.nl/$\sim$jarle/SDSS}. We correct the originally underestimated H$\beta$ fluxes in the catalog as outlined in \cite{groves12}, and apply the scalings to the line flux errors recommended by \cite{juneau14}. As in \cite{tremonti04} and \cite{kewley08} we apply several cuts to the original MPA/JHU catalog to build a representative sample of local star forming systems while minimizing aperture effects due to the 3'' SDSS fiber size. 

We only consider objects with reliable stellar masses. We define this as galaxies with a 68\% stellar mass confidence interval (i.e. 1$\sigma$ for a Gaussian PDF) smaller than 0.2~dex. We restrict our analysis to galaxies in the $10^{8.7}$~M$_{\odot}<M_*<10^{11.1}$~M$_{\odot}$ range to ensure good enough statistics for building the MZR (see Section \ref{sec:results-highmass}). We further require a redshift in the $0.04<z<0.25$ range, and a SDSS fiber radius ($r_f=3"$) larger than half of the g-band Petrosian effective radius ($r_e$, {\tt PETROR50}) measured by the SDSS imaging pipeline. The latter cuts ensure a significant fraction of the total nebular emission falls within the SDSS fiber. We also require $S/N\ge 8$ in the [OII]$\lambda$3727, H$\beta$, [OIII]$\lambda$5007, H$\alpha$, [NII]$\lambda6584$, and [SII]$\lambda6717$ lines used in the metallicity measurements. Finally, all objects are required to fall in the purely ``star forming'' region of the [NII] BPT diagram \citep{baldwin81}. We define this region using the empirical criteria given in Equation 4 of \cite{kewley06}.

All these selections translate into a final sample of 43,306 objects. Figure \ref{fig:m-sfr} shows how the objects in this sample follow a $M*$-SFR relation typical of local star forming galaxies. This sample selection is similar but not identical to the ones used in \cite{tremonti04} and \cite{kewley08}. In Section \ref{sec:discussion-test-samples} we show that adopting the selection cuts used in those studies instead of the ones described here produces indistinguishable results regarding the shape of the MZR.

\subsubsection{Local Volume Dwarf Galaxy Sample}

To measure the MZR in the low stellar mass regime we use the sample of local volume (LV) dwarf galaxies with $10^{5.9}$~M$_{\odot} \le M_* \le 10^{9.3}$~M$_{\odot}$ from \cite{lee06}. These are 27 nearby star forming dwarfs ($D\le5$ Mpc) with reliable stellar masses based on {\it Spitzer}-IRAC photometry and high quality nebular abundance estimations in the literature. We do not adopt the literature values for the gas-phase metallicity of these objects but rather compile the [OII]$\lambda$3727, H$\beta$, [OIII]$\lambda5007$, H$\alpha$, [NII]$\lambda6584$, and [SII]$\lambda6717$ line fluxes published in the original references \citep[see Table 1 in][]{lee06} and re-compute the metallicity using {\tt IZI}.

For 21 of these 27 objects we are able to compile enough strong line emission line fluxes as to allow reliable metallicity estimations. Five objects (DDO53, DDO165, HoI, HoII, HoIX, and NGC3738) either lack a reported flux for both [OII] and [SII] making the ionization parameter determination ambiguous, or have line ratios not corrected for extinction instead of fluxes reported in the original references. For this sample we adopt the stellar masses in Table 1 of \cite{lee06}. Red points in Figure \ref{fig:m-sfr} show 18 of these 21 galaxies which have available UV SFR measurements from GALEX far-UV photometry obtained as part of the 11HUGS survey \citep{lee09}. The Local Volume Dwarf sample closely follows the extrapolation towards low masses of the SDSS high mass sample $M*$-SFR relation (dashed line in Figure \ref{fig:m-sfr}).

\subsubsection{SDSS Low Mass Sample}

The LV dwarfs sample of \cite{lee06} is relatively small and restricted to a particular environment in the Universe (the Local Group and its vicinity). In order to overcome these limitations we also include a sample of low mass, low metallicity galaxies selected from the SDSS DR3 \citep{abazajian05} "main galaxy" sample by \cite{izotov06}. The main selections in this sample are the requirement of a 1$\sigma$ detection of the [OIII]$\lambda$4363 auroral line, and an H$\beta$ emission line flux brighter than $10^{-14}$~erg~s$^{-1}$~cm$^{-2}$. Further cuts based on the [OII]$\lambda$3727, HeII$\lambda$4686, H$\beta$, [OIII]$\lambda$5007, and [NII]$\lambda$6584 lines are used to remove AGNs, and galaxies dominated by low excitation regions for which the [OIII]$\lambda$4363 auroral line might be affected by heating mechanisms not associated with stellar radiation, as described in \cite{izotov06}. While this sample was originally selected from SDSS DR3, in our analysis we use the DR7 corrected emission line fluxes and errors for the same transitions used for the high mass SDSS sample (see Section \ref{sec:methods-samples-sdss}).

We use this sample after applying a redshift cut of $z>0.02$ and a stellar mass cut at $10^{7.5}$~M$_{\odot} \le M_* \le 10^{10}$~M$_{\odot}$. The former ensures the [OII]$\lambda$3727 line falls within the SDSS wavelength range, while the latter ensures good enough statistics to measure the MZR. After these cuts we obtain a sample of 129 galaxies that extend the range of the main SDSS sample described above by one order of magnitude towards lower stellar masses. The non-trivial sample selection of \cite{izotov06} has the potential to introduce biases in the derived MZR. Particularly, the need for an auroral line detection requires these galaxies to have bright emission lines, and therefore high SFRs (blue dots in Figure \ref{fig:m-sfr}). It also biases the sample against high metallicity galaxies with low electron temperatures. We discuss these effects in Section \ref{sec:results-lowmass} where we find the SFR bias to produce a negligible impact on the MZR shape, and the abundance bias to significantly affect the MZR only at the high mass end of the sample.

\subsubsection{DEEP2 Low Mass Sample}

We also include in our analysis a sample of low mass $z<0.38$ galaxies selected by \cite{zahid12} using data from the DEEP2 survey \citep{davis03}. DEEP2 probes significantly fainter objetcs than SDSS ($R<24.1$ versus $r'<17.77$), therefore probing lower stellar masses. DEEP2 covers a more limited wavelength range (6500\AA-9100\AA) than SDSS so only the H$\alpha$, [NII]$\lambda\lambda$6548,6584, and [SII]$\lambda\lambda$6717,6731 transitions are available for these galaxies\footnote{H. J. Zahid, private communication.}. The main selections for this sample include a $S/N>3$ requirement for H$\alpha$, a 1$\sigma$ uncertainty in stellar mass smaller than $0.3$ dex, and a $\chi^2<2$ cut on the quality of the fit used to measure the emission lines. Further cuts on galaxy color and the [NII]/H$\alpha$ ratio are used to remove AGNs from the sample. The details of the selection are described in \cite{zahid12}. We further require that the fit yields a measurements of the [SII] doublet, and a stellar mass in the $10^{7.25}$~M$_{\odot} \le M_* \le 10^{10}$~M$_{\odot}$ range to ensure enough statistics to measure the MZR. 

The adopted selection keeps the bulk of the objects in the original \cite{zahid12} sample (624 out of 777). Similarly the SDSS Low Mass sample, the cuts used to build this sample translate into a bias towards high SFRs at fixed stellar mass (orange dots in Figure \ref{fig:m-sfr}), but not in a strong abundance bias. The SFR bias is found to not impact significantly the shape of the MZR (see Section \ref{sec:results-lowmass}).

\subsection{Measuring the Oxygen Abundance PDF with {\tt IZI}}

In \cite{blanc15} we describe a new method to derive the oxygen abundance \mbox{($12+\log{\rm{O/H}}$)} and ionization parameter ($q$) of ionized nebulae using Bayesian statistics, and we present a software tool called {\tt IZI} that implements this approach. In brief, {\tt IZI} calculates the joint and marginalized posterior PDFs for \mbox{$12+\log(\rm{O/H})$} and $\log{(q)}$ given a set of observed emission line fluxes (and their associated errors) and an input photo-ionization model. The method allows the use of arbitrary sets of SELs, the inclusion of flux upper limits, and it provides a self-consistent way of determining the physical conditions of ionized nebulae that is not tied to the arbitrary choice of a particular SEL diagnostic, while using all available information. 

Unlike theoretically calibrated SEL diagnostics {\tt IZI} is flexible and not tied to a particular photo-ionization model. \cite{blanc15} provides a detailed description of the algorithm and an exhaustive comparison to other methods in the literature. In particular \cite{blanc15} find that {\tt IZI} in combination with
the MAPPINGS-III photoionization models of \cite{levesque10} performs very well when comparing the derived abundances with oxygen RL abundances in a sample of 22 local HII regions for which such measurements are available. The method shows a $\lesssim 0.1$ dex systematic offset and a $\sim 0.1$ dex
scatter with respect to RL abundances. Therefore we adopt the \cite{levesque10} models and use {\tt IZI} to measure the oxygen abundance for all the galaxies in the samples described in Section \ref{sec:methods-samples}. 

For galaxies in both SDSS samples we calculate the $12+\log{(\rm{O/H})}$ PDF (marginalized over the ionization parameter) for each individual object using the fiber flux of all the SEL transitions mentioned in Section \ref{sec:methods-samples-sdss}. By default {\tt IZI} assumes uniform maximum ignorance priors in $\log{(q)}$ and $12+\log{(\rm{O/H})}$.
Maximum ignorance is also assumed for galaxies in the DEEP2 sample, but only the the H$\alpha$, [NII]$\lambda\lambda$6548,6584, and [SII]$\lambda\lambda$6717,6731 fluxes are used in the fit. In Section \ref{sec:results-lowmass} we will show that the DEEP2 and SDSS MZRs agree well within the errors. This highlights a fundamental advantage of {\tt IZI}: the ability to measure self-consistent abundances with realistic error-bars while using different sets of emission lines.

For the Local Volume low mass sample we adopt an extra prior of $12+\log{(\rm{O/H})}\le8.5$ (equivalent to the common assumption that these galaxies are in the low metallicity branch of $R23$). This prior helps improve the fits for a few objects but removing it does not significantly affect our results. For this sample SEL fluxes are usually reported for several individual HII regions in each object (5 on average but up to 17 in some cases). We obtain the best-fit metallicity and 68\% confidence intervals for each individual HII region \citep[adopting the joint-mode best fit parameters, see][]{blanc15}, and then average the results to obtain the metallicity PDF of each galaxy. These values are reported in Table \ref{tbl-1}.

In 19 of the 21 LV dwarf galaxies the dispersion between individual HII region abundances is smaller or equal than the average uncertainty in the measurements. This highlights the known fact that the ISM of most dwarf galaxies is largely chemically homogeneous \citep[e.g.][]{croxall09}. Two dwarf galaxies present significant dispersion in the abundance of their individual HII regions. In WLM one HII region (HM 19) shows a significantly lower abundance (-0.4~dex) than the rest. This was already noticed by \cite{lee05} and it could be associated with star formation triggered by recent accretion of low metallicity gas. The other galaxy showing large dispersion (0.26~dex) is Sextans B, which has been reported to present significant abundance inhomogeneities in its ISM \citep{kniazev05}. None of these two galaxies show significant deviations from the rest of the sample in the MZR so we include them in the analysis.

\section{Results}\label{sec:results}

\subsection{The SDSS High Mass Sample MZR}
\label{sec:results-highmass}

Using the SDSS high mass sample, we build the MZR by summing the oxygen abundance PDFs of individual galaxies in \mbox{0.05 dex} bins in stellar mass, and then renormalizing the distributions by the total number of galaxies in each bin. We restrict the analysis to the \mbox{$10^{8.7}$~M$_{\odot}<M_*<10^{11.1}$~M$_{\odot}$} range to ensure a sufficient number of galaxies per bin. On average there are $\simeq900$ objects per stellar mass bin, and the lowest number of objects is observed in the lowest mass bin which contains 15 galaxies.

\begin{figure}[t!]
\begin{center}
\epsscale{1.15}
\plotone{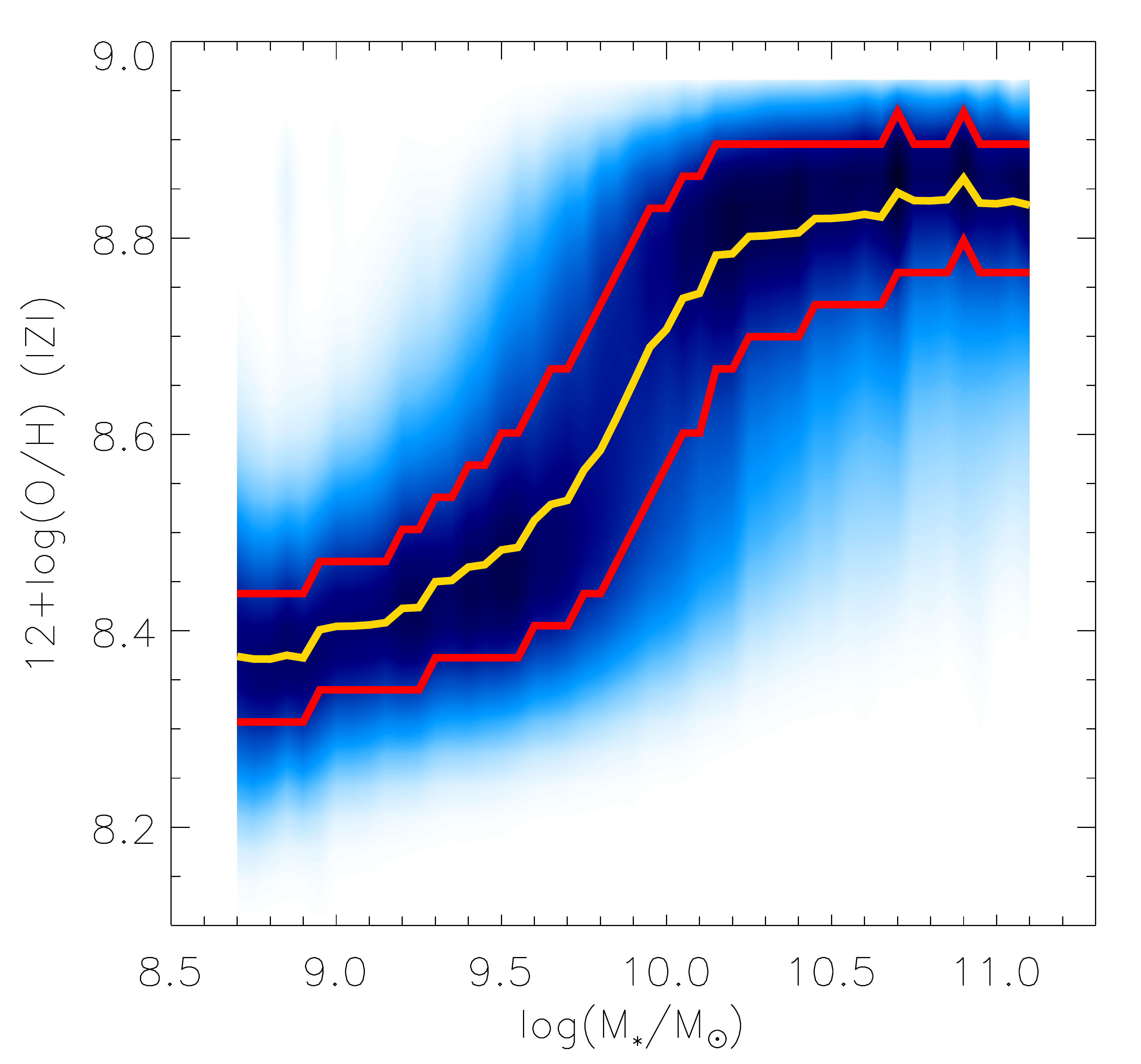}
\caption{Gas-phase MZR for \mbox{$\sim43,000$} galaxies
  in SDSS DR7. The MZR is constructed by summing the marginalized PDFs
  for the oxygen abundance of individual galaxies in bins of 0.05 dex
  in stellar mass (blue). The mean of the total PDF and the bounds of the 68\%
  confidence interval are shown by the yellow and red solid lines
  respectively.}
\label{fig:mzr-sdss}
\end{center}
\end{figure}

Figure \ref{fig:mzr-sdss} presents the resulting MZR measured using this method. The mean and $\pm1\sigma$ envelope (68\% confidence interval) for the oxygen abundance as a function of stellar mass are shown by the yellow and red solid lines respectively. These curves are also reported in Table \ref{tbl-2}. We observe a tight relation between mass and metallicity with a $1\sigma$ confidence interval of $\simeq 0.1$~dex around the mean. This is consistent with the typical scatter in the MZR seen by other authors using the same dataset \citep[e.g.][]{tremonti04}. The median $1\sigma$ confidence interval for the metallicity of individual galaxies using our method is also $\simeq0.1$ dex, implying that the intrinsic scatter in the relation over this stellar mass range is lower than this value. 

We observe different behavior in the MZR over three distinct ranges in stellar mass. At high masses ($M_*>10^{10.5}$~M$_{\odot}$) the MZR flattens to a constant value of about $12+\log{\rm{O/H}}\simeq8.8$. This behavior has been observed in previous studies \citep[e.g.][]{tremonti04, kewley08, zahid14a} and corresponds to a saturation of the gas-phase metallicity when it approaches the value of the effective metal production yield. As the metallicity of the ISM increases, the amount of metals locked into long lived low mass stars and stellar remnants also increases, eventually balancing the addition of new metals from newly formed stars \citep{zahid14a}. For stellar masses below this saturation limit we observe that the MZR behaves differently over two distinct regimes in stellar mass. 
In the \mbox{$10^{9.5}$~M$_{\odot}<M_*<10^{10.5}$~M$_{\odot}$} range we observe a steep increase in $Z$ as a function of $M_*$, while a significantly shallower relation is seen in the \mbox{$10^{8.7}$~M$_{\odot}<M_*<10^{9.5}$~M$_{\odot}$} range.

We fit three separate power-law functions ($Z\propto M_*^\alpha$) to the observed relation over these three stellar mass ranges. These fits yield slopes of $\alpha=0.14\pm0.08$ for the \mbox{$10^{8.7}$~M$_{\odot}<M_*<10^{9.5}$~M$_{\odot}$} range, $\alpha=0.37\pm0.08$ for the
\mbox{$10^{9.5}$~M$_{\odot}<M_*<10^{10.5}$~M$_{\odot}$} range, and $\alpha=0.03\pm0.11$ for \mbox{$M_*>10^{10.5}$~M$_{\odot}$}. Therefore, there appears to exist a characteristic transition mass scale at $M_*\simeq10^{9.5}$~M$_{\odot}$ past which the ISM of galaxies suffers a high level of chemical enrichment as the stellar mass increases.

\begin{figure}[t!]
\begin{center}
\epsscale{1.15}
\plotone{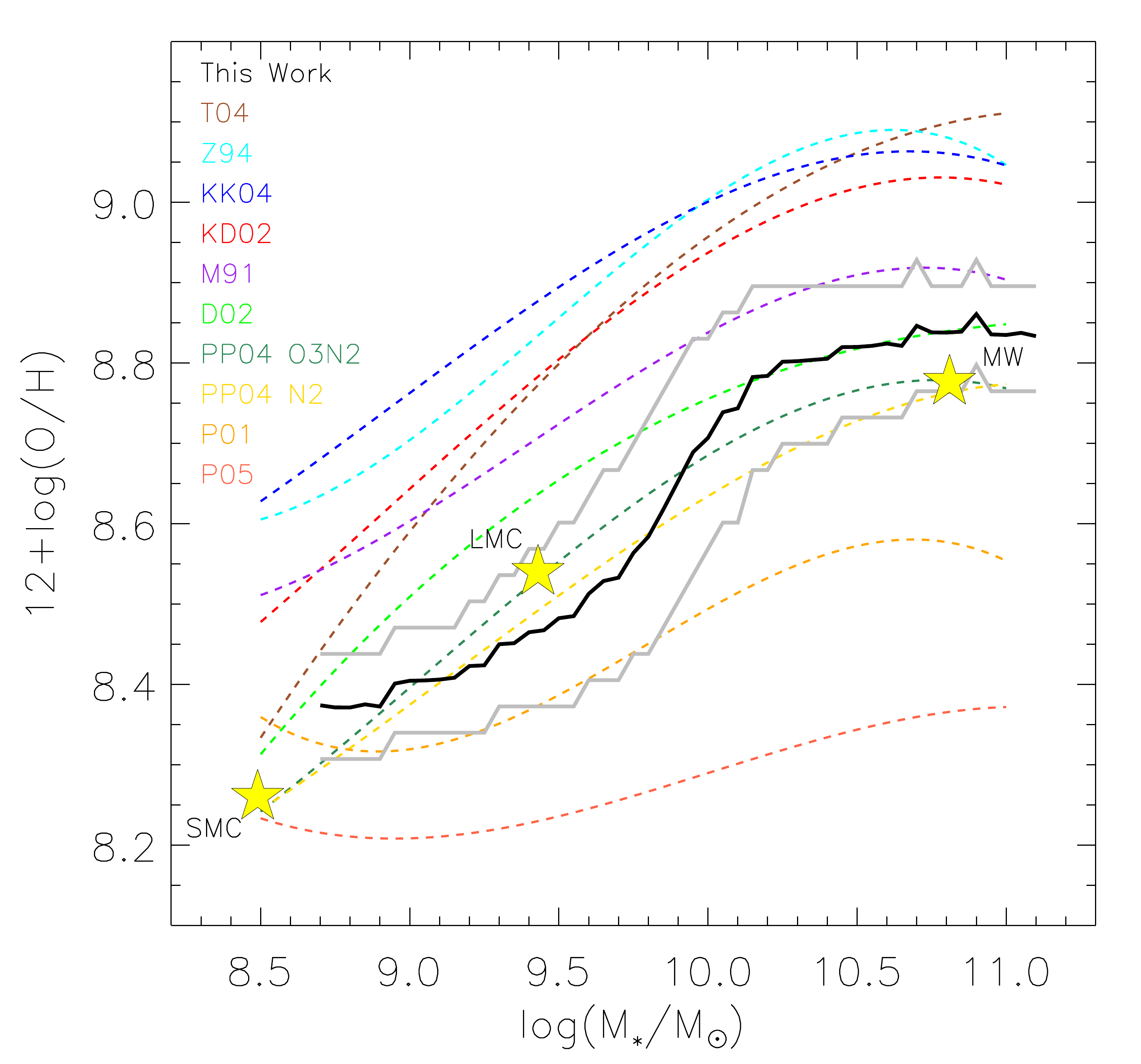}
\caption{Comparison between the MZR measured with {\tt IZI} and the
  photo-ionization models of \cite{levesque10}, and the MZR measured
  using ten different SEL diagnostics by \citealt{kewley08} (dashed
  color lines). Our measurement with its 68\% confidence interval
  is shown by the black and grey solid lines. The positions of the
  MW, LMC, and SMC galaxies in the MZR from high quality abundance
  determinations are marked by the yellow stars.}
\label{fig:mzr-comp}
\end{center}
\end{figure}

This behavior is qualitatively different from what has been typically reported in previous studies of the MZR of star forming galaxies where a
single power-law shape is reported at low and intermediate masses, followed by a plateau in metallicity towards the high mass end \citep[e.g.][]{tremonti04, mannucci10, zahid12, zahid14b, andrews13}. Since several of these measurements are based on similar subsets of the same dataset used here, the differences in observed behavior are most likely associated with the methods used to measure the oxygen abundance in different studies (see Section \ref{sec:discussion-test-abundances}).

Figure \ref{fig:mzr-comp} presents a comparison between our measurement of the MZR using {\tt IZI}, and a series of measurements of the SDSS MZR from \cite{kewley08}. Different colors represent different SEL abundance diagnostics from the literature. Also shown for reference are the position in the MZR of the Milky Way (MW), the Large Magellanic Cloud (LMC), and the Small Magellanic Cloud (SMC). For the Magellanic system we adopt the stellar masses reported in \cite{vandermarel09} ($M_*=10^{8.49}$~M$_{\odot}$
and $10^{9.43}$~M$_{\odot}$ for the SMC and LMC respectively), and for the MW we adopt the
$M_*=10^{10.81}$~M$_{\odot}$ value measured by \cite{mcmillan11}. The oxygen abundances are taken from \cite{penaguerrero12}, \cite{peimbert03}, and \cite{garciarojas04} for the SMC ($12+\log{(\rm{O/H})}=8.26$), LMC ($12+\log{(\rm{O/H})}=8.54$), and MW ($12+\log{(\rm{O/H})}=8.78$) respectively. All these values come from RL abundances (LMC and MW) or direct method abundances corrected for electron temperature inhomogeneities (SMC). For the SMC we have averaged the abundances measured for two HII regions (NGC~456 and NGC~460) by \cite{penaguerrero12}. For the MW we have evaluated the abundance gradient of \cite{garciarojas04} at a galactocentric radius of 1.1 kpc, corresponding to half of the SDSS fiber radius at the median redshift of the high mass sample ($\langle z \rangle=0.077$). 

\begin{figure*}[t]
\begin{center}
\epsscale{1.15}
\plotone{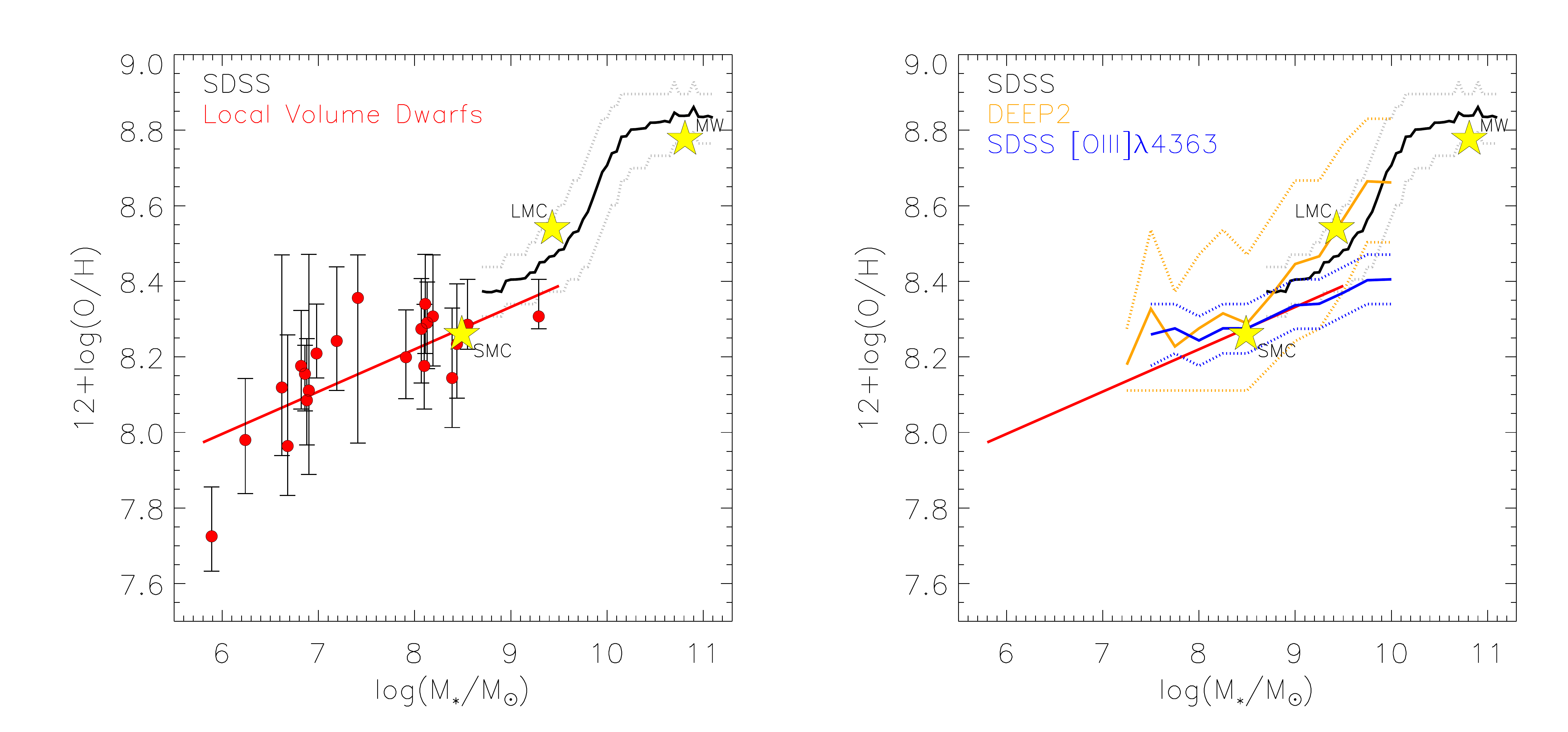}
\caption{MZR for the three dwarf galaxy samples analyzed in this work together with the SDSS massive galaxy sample MZR (black solid line with 1$\sigma$ confidence intervals in grey). The left panel shows the Local Volume dwarf galaxy sample of \cite{lee06} (red circles with errorbars) and the best linear fit to this sample alone (solid red line). The right panel shows the MZR for the low mass SDSS sample of \citealt{izotov06} (in blue) and the DEEP2 low mass sample of \citealt{zahid12} (in orange). The position of the MW, LMC, and SMC galaxies in the MZR from high quality abundance
  determinations are marked by the yellow stars.}
\label{fig:mzr-dwarfs}
\end{center}
\end{figure*}

As widely discussed in the literature \citep[e.g.][]{kewley08, moustakas10, lopez-sanchez12, blanc15} large offsets in the absolute abundance scale are seen between different SEL diagnostics. A general trend often seen is that diagnostics that are empirically calibrated against the direct method result in abundances that are $\sim 0.2-0.3$~dex lower than those calibrated against theoretical photo-ionization models. While this holds true on average \cite{blanc15} also find that large offsets can be seen between abundances calculated using different photo-ionization models in the literature. In that work we show that the models of \cite{levesque10}, adopted here, provide an excellent match to the significantly more reliable RL abundances of local HII regions over the $8.0<12+\log{(\rm{O/H})}<8.8$ range sampled by the 22 nearby bright regions for which oxygen RL measurements are available in the literature\footnote{As seen in Figure 9 of \cite{blanc15}, only one HII region (NGC2363) at $12+\log{(\rm{O/H})}<8.3$ has its RL abundance measured and reported in the literature. While the RL and IZI abundances for this region agree well within the error bars, one could say that the agreement has been ``robustly'' established in the $8.3<12+\log{(\rm{O/H})}<8.8$ range.}. This is in line with the fact that the MZR measured here is consistent with the high quality abundance measurements for the MW, the LMC, and the SMC quoted above. These measurements are also consistent with the $N2$ and $O3N2$ SEL diagnostics.

\subsection{The MZR in the Low Mass Regime}
\label{sec:results-lowmass}

Figure \ref{fig:mzr-dwarfs} presents the MZR measured for the massive galaxies in SDSS together with measurements for the three low mass galaxy samples described in Section \ref{sec:methods-samples}: the LV dwarf galaxy sample of \cite{lee06}, the low mass, low metallicity sample of SDSS galaxies from \cite{izotov06}, and a sample of low mass systems selected from the DEEP2 survey by \cite{zahid12}. The oxygen abundances of all galaxies have been remeasured using {\tt IZI} so the results for all samples should be self-consistent. 

For the Local Volume dwarfs we plot individual galaxies (Table \ref{tbl-1}) in the left panel of Figure \ref{fig:mzr-dwarfs}. Error bars correspond to the 68\% confidence intervals of the marginalized metallicity PDF. This sample covers the $10^{5.9}$~M$_{\odot}<M_*<10^{9.3}$~M$_{\odot}$ range in stellar mass and its MZR appears as a smooth continuation of the relation seen at the low mass end of the SDSS high mass sample. A power law fit to the MZR of these 21 objects yields a slope of $\alpha=0.11\pm0.03$, which is in agreement with the $\alpha=0.14\pm0.08$ best-fit slope of the low mass ($10^{8.7}$~M$_{\odot}<M_*<10^{9.5}$~M$_{\odot}$) end of the SDSS high mass MZR. A $\simeq0.07$ dex normalization offset between the two relations is seen in the overlap region around $10^{9}$~M$_{\odot}$. This offset is within the 1$\sigma$ uncertainty in both measurements, and is consistent in magnitude and direction with the expected bias in the SDSS abundances introduced by metallicity gradients and fiber aperture effects as shown in Section \ref{sec:discussion-test-aperture}. 

\begin{figure*}[t]
\begin{center}
\epsscale{1.15}
\plotone{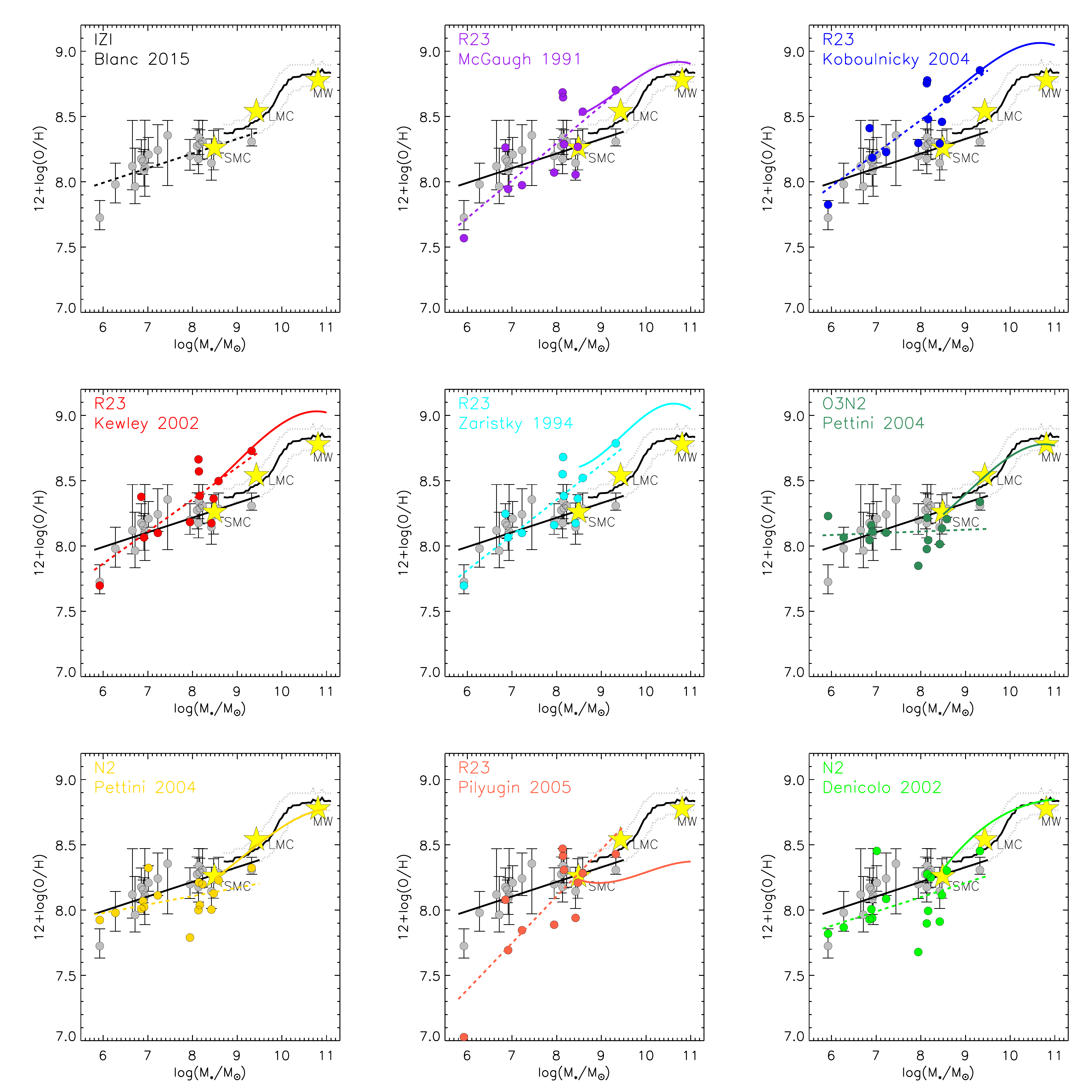}
\caption{Same as the left panel in Figure \ref{fig:mzr-dwarfs} but overlaying the MZR calculated using eight different SEL diagnostics compiled in the Appendix of \cite{kewley08}. Results based on {\tt IZI} are shown in black/grey. Colored circles show the average metallicity of HII regions in Local Volume dwarf galaxies using each diagnostic, with the best fit linear relation showed as a dashed line. The solid lines show the same SDSS MZRs from \cite{kewley08} shown in Figure \ref{fig:mzr-comp}.}
\label{fig:mzr-dwarfs-lit}
\end{center}
\end{figure*}

The fact that the MZR for these two independent datasets agrees both in slope and normalization is encouraging. Integrated measurements of ionized emission over large areas across galaxies (as in SDSS fiber spectroscopy) can in principle introduce biases in the measured oxygen abundance with respect to what would be otherwise measured from spectroscopy of individual HII regions (as in the Local Volume Dwarf sample). Beyond the obvious radial sampling effect introduced by metallicity gradients, which we explore in detail in \ref{sec:discussion-test-aperture}, the mixing of ensembles of HII regions with different physical properties and contamination from diffuse ionized gas \citep[DIG,][]{reynolds73, madsen06, haffner09, zhang17} has the potential to bias integrated line flux measurements. In Section \ref{sec:discussion-test-dig} we discuss these effects in detail and conclude that their impact on our results is expected to be small. Metallicities measured for low mass galaxies using integrated spectroscopic measurements over the SDSS fiber, do not seem to suffer significant systematic biases, and yield consistent results with targeted slit spectroscopy of individual HII regions.

The right panel of Figure 3 compares the high mass sample MZR with those measured in the low mass SDSS and DEEP2 samples. The larger sizes of these samples compared to that of \cite{lee06} allow us to use the same methodology adopted to build the SDSS high mass MZR. This is, summing the metallicity PDF of individual objects in stellar mass bins. These samples cover the $10^{7.25}$~M$_{\odot}<M_*<10^{10}$~M$_{\odot}$ range, and in both cases a stellar mass bin of 0.25 dex is used when building the MZR to ensure a minimum of 3 objects per bin. The average number of galaxies per stellar mass bin are 12 and 21 for the \cite{izotov06} and \cite{zahid12} samples respectively. Following \cite{zahid12} we scaled up the stellar masses derived in that work by a 0.2~dex offset to make them consistent with the SDSS MPA/JHU stellar masses.

Below $10^9$~M$_{\odot}$ the MZR for all three low mass samples agree well within the errors and have slopes that are consistent with each other and with the low mass end of the high mass SDSS sample. In this range we measure $\alpha=0.06\pm0.05$ and $\alpha=0.12\pm0.09$ for the low mass SDSS and DEEP2 samples respectively. For higher stellar masses the DEEP2 sample agrees well within the errors with the high mass SDSS sample, but the \cite{izotov06} is biased towards lower abundances. This bias is expected from the requirement of a [OIII]$\lambda$4363 detection in this sample. This selection rejects high metallicity systems in which the auroral line becomes too faint to be detected due to enhanced metal cooling and low electron temperatures. 

The significant offset towards high SFR at fixed stellar mass seen in Figure \ref{fig:m-sfr} for the DEEP2 and low mass SDSS samples does not translate into a significant offset in the MZR. This is in line with recent results finding either a lack, or a very mild dependence of metallicity with SFR at fixed stellar mass \citep{hughes13, sanchez13, salim14, barrera17}, at odds with the proposed existence of a strong "fundamental" mass-metallicity-SFR relation  for galaxies \citep{mannucci10, lara-lopez10}. We discuss this issue in detail in \ref{sec:discussion-tests-sfr}.

In Figure \ref{fig:mzr-dwarfs-lit} we compare the MZR measured with {\tt IZI} for Local Volume dwarfs and the high mass SDSS sample, against the results from using eight different SEL diagnostics in the literature compiled by \cite{kewley08}. These are the same methods used in the comparison shown in Figure \ref{fig:mzr-comp}, which we now apply to HII regions in the Local Volume dwarf sample galaxies. Unlike {\tt IZI} which works with an arbitrary set of available lines, these methods can only be applied when all the lines in the adopted diagnostic ratios are available, which is not the case for all galaxies. We fit linear relations to the dwarf MZR measured with each method. For simplicity we adopt a uniform 0.15~dex uncertainty for each galaxy when doing this fit. This error is representative of the scatter typically seen in SEL diagnostics calibrations. Since the \cite{lee06} study is based on direct method abundances derived from spectra deep enough to detect auroral lines, the signal to noise in the SEL is large enough that the formal errors in the line fluxes can be neglected.

For most $R23$ methods the MZR for dwarf and high mass galaxies joins smoothly and shows a single power-law beheaviour towards low masses. The apparent break in the MZR for the \cite{pilyugin05} $R23$ calibration is largely driven by the extremely low metallicity that this method implies for the lowest mass galaxy in the sample (Leo A). Further inspection shows that this is driven by two extreme low ionization HII region in this system for which the ionization parameter correction in the method drives the oxygen abundance down by $~0.7$~dex with respect to the average of the other three HII regions which show normal ionization conditions. This reflects the limitations of empirically calibrated methods, which can only produce reliable results for regions in the same parameter space as those in the adopted calibration samples. Interestingly, both $N2$ and $O3N2$ calibrations produce a MZR that presents a similar behavior as the one we observe with {\tt IZI}, with a slope that flattens towards low stellar masses. Although these calibrations might no be as reliable in the low abundance regime as dicussed in Section \ref{sec:discussion-test-abundances}. This confirms the conclusion of \cite{kewley08} that the choice of abundance diagnostics can have a significant effect on the observed shape of the MZR, and shows that other SEL diagnostics besides {\tt IZI} can also produce a MZR that changes slopes towards the low mass regime.

\section{Analysis of potential systematic errors}\label{sec:systematics}

In this section we  explore the impact of several potential sources of systematic error in our measurement of the MZR. We investigate the effects of sample selection, spectroscopic aperture effects, a changing N/O abundance in low mass galaxies, the statistical methods used to characterize the metallicity PDFs and build the MZR, a potential secondary dependence of the MZR with the SFR, and the impact of luminosity weighted averaging and DIG contamination. We also explore the systematic differences between {\tt IZI} abundances and those computed using the direct method.

\subsection{Systematics Associated with Sample Selection and Methodology}
\label{sec:discussion-tests}

This analysis is largely focused on the SDSS high mass sample described in Section \ref{sec:methods-samples-sdss}, as we have control of all the selection parameters for this sample. Here we focus on the methods used to select the sample, measure line fluxes, and construct the MZR. The next section focuses on the diagnostics used to measure the oxygen abundance. 

\subsubsection{Adopted S/N Threshold}

When selecting the SDSS high mass sample we used a $S/N\ge8$ cut on all SELs used in the abundance determination. This $S/N$ threshold is the same one used by \cite{kewley08}, but other authors have applied different cuts (anywhere from $S/N=3$ to 25) to different subsets of lines when studying the MZR of local star forming galaxies \citep[e.g.][]{tremonti04, mannucci10, foster12}. Here we explore the effect of this $S/N$ threshold on the observed MZR measured with {\tt IZI}. 

The left and center panels in the top row of Figure \ref{fig:mzr-tests} show the MZR for samples selected in the same way as the SDSS high mass sample except for the use of lower $S/N$ thresholds on all lines ($S/N\ge5$ and $S/N\ge3$ respectively). The shape of the MZR is largely insensitive to differences in the $S/N$ threshold, with $<1\sigma$ deviations from the $S/N\ge8$ case. Lowering the $S/N$ threshold induces a small systematic offset towards higher abundances at the high mass end. This is consistent with the results of \cite{foster12} who find that more stringent $S/N$ cuts on the $[OIII]\lambda 5007$ line preferentially reject high abundance, high stellar mass, systems in which cooling is increasingly dominated by the infrared oxygen transitions and the optical [OIII] lines are weak. The presence of a characteristic transition mass scale at $\sim10^{9.5}$~M$_{\odot}$ in the MZR is not affected by the $S/N$ cuts used to construct the sample.

\subsubsection{SDSS Fiber Aperture Effects}
\label{sec:discussion-test-aperture}

It is a well established result that massive star forming disk galaxies in the local universe present, on average, negative radial gradients in their
oxygen abundance \citep{searle71, mccall85, vilacostas92, zaritsky94, sanchez14, kaplan16}. The fixed $3''$ angular diameter of the SDSS spectrograph
optical fibers implies that a different fraction of the star forming disk will contribute to the measured emission line fluxes depending on the distance and physical size of the target. These differences in sampling can translate into biases in the measured metallicities if gradients are present. If the fiber only covers the very central part of the disk the measured metallicity will most likely be higher than if the fiber samples the disk out to larger radii. We try to minimize this effect during the selection of the SDSS high mass sample by enforcing that the fiber radius covers a significant fraction of
the galaxies effective radii ($r_f\ge 0.5 r_e$). Nevertheless, residual aperture effects might still be present, and it is important to evaluate how they affect the shape of the observed MZR.

To do so, we use a statistical approach to quantify a correction to the metallicity PDF of individual galaxies, based on their
$r_f/r_e$ ratio, and the average metallicity gradient observed for star forming galaxies in the local universe. We then construct the MZR using these corrected PDFs and compare the results to the uncorrected MZR presented in Figure \ref{fig:mzr-sdss}.

\cite{sanchez14} used IFU data for 306 nearby star forming galaxies in the CALIFA survey \citep{sanchez12} to measure the oxygen abundance in more than 7000 associations of HII regions. They find that, on average, local disk galaxies follow a roughly universal abundance gradient with a characteristic value of $\alpha_{\rm{O/H}}=-0.1$ dex $r_e^{-1}$. We use this result to calculate our aperture corrections assuming all the objects in the SDSS sample follow this average gradient. Under these assumptions the observed luminosity weighted abundance inside the fiber is given by:

\begin{equation}
\log{(\rm{O/H})_f}=\frac{\int_0^{r_f}I(r)\left[\log{(\rm{O/H})_e}+\alpha_{\rm{O/H}}\left(\frac{r}{r_e}-1\right)\right]rdr}{\int_0^{r_f}I(r)rd}.
\end{equation}

\noindent
where $\log{(\rm{O/H})_e}$ is the oxygen abundance at the effective radius, and $I(r)$ the nebular emission surface brightness profile. Solving this equation allows one to calculate a statistical correction that we apply separately for each object. 

We adopt the SDSS g-band Petrosian effective radius as $r_e$, the $\alpha_{\rm{O/H}}$ value from \cite{sanchez14}, and a $3''$ angular diameter for the SDSS fiber. We assume two different functional forms for the surface brightness profile: flat and exponentially declining, and use Equation 1 to calculate the offset between the measured oxygen abundance in the SDSS fiber and the characteristic abundance at the effective radius. In both cases we find these corrections to be small compared to the uncertainty in the metallicity. For the flat surface brightness profile we find corrections in the $-0.07<\Delta\log{(\rm{O/H})}<0$ range with an average correction of $-0.05$~dex. For the exponential surface brightness profile we find corrections in the $-0.1<\Delta\log{(\rm{O/H})}<0.03$ range with an average correction of $-0.08$~dex. 

In both cases there is no apparent correlation between galaxy stellar mass and the magnitude of the abundance aperture correction. Therefore, the shape of the MZR is not affected by these corrections beyond a small offset towards lower abundances. This can be seen in the right panel of the top row of Figure \ref{fig:mzr-tests}, where the MZR measured after applying aperture corrections for a flat surface brightness profile, is compared with the uncorrected measurement. The result remains unaltered when assuming an exponential profile, except for a -0.03~dex offset towards lower abundances. We conclude that the observed characteristic mass scale in the MZR does not arise as a result of fiber aperture effects in the metallicities. This conclusion is also supported by the fact that the shape of the MZR in the low mass regime of the SDSS sample agrees with that measured for the LV dwarf sample, in which abundances are measured using slit spectroscopy of individual HII regions.



\subsubsection{Effects Introduced by the N/O Abundance Ratio}
\label{sec:discussion-test-n/o}

The N/O abundance ratio in the ISM of galaxies is not constant as a function of metallicity but instead shows a transition between a primary and secondary nitrogen production regimes \citep{vanzee06}. While the photo-ionization models used by {\tt IZI} include an average N/O vs. O/H relation that models this effect, a systematic deviation in our sample from this average relation could in principle induce a bias in the MZR (see discussion in Section \ref{sec:discussion-literature}). So it is interesting to test the impact that the N/O abundance could have on the MZR measured by {\tt IZI}. To do so we conduct a simple test in which we remove the [NII] lines from the set of transitions given as input to {\tt IZI} for the calculation of the metallicity PDFs. The ratio of the [NII] doublet to other SELs has been shown to depend strongly on the assumed N/O abundance \citep{perez-montero09, perez-montero14}, therefore removing this transition form the fit largely removes the dependence of the output PDFs on the N/O ratio.

\begin{figure*}[t!]
\begin{center}
\epsscale{1.15}
\plotone{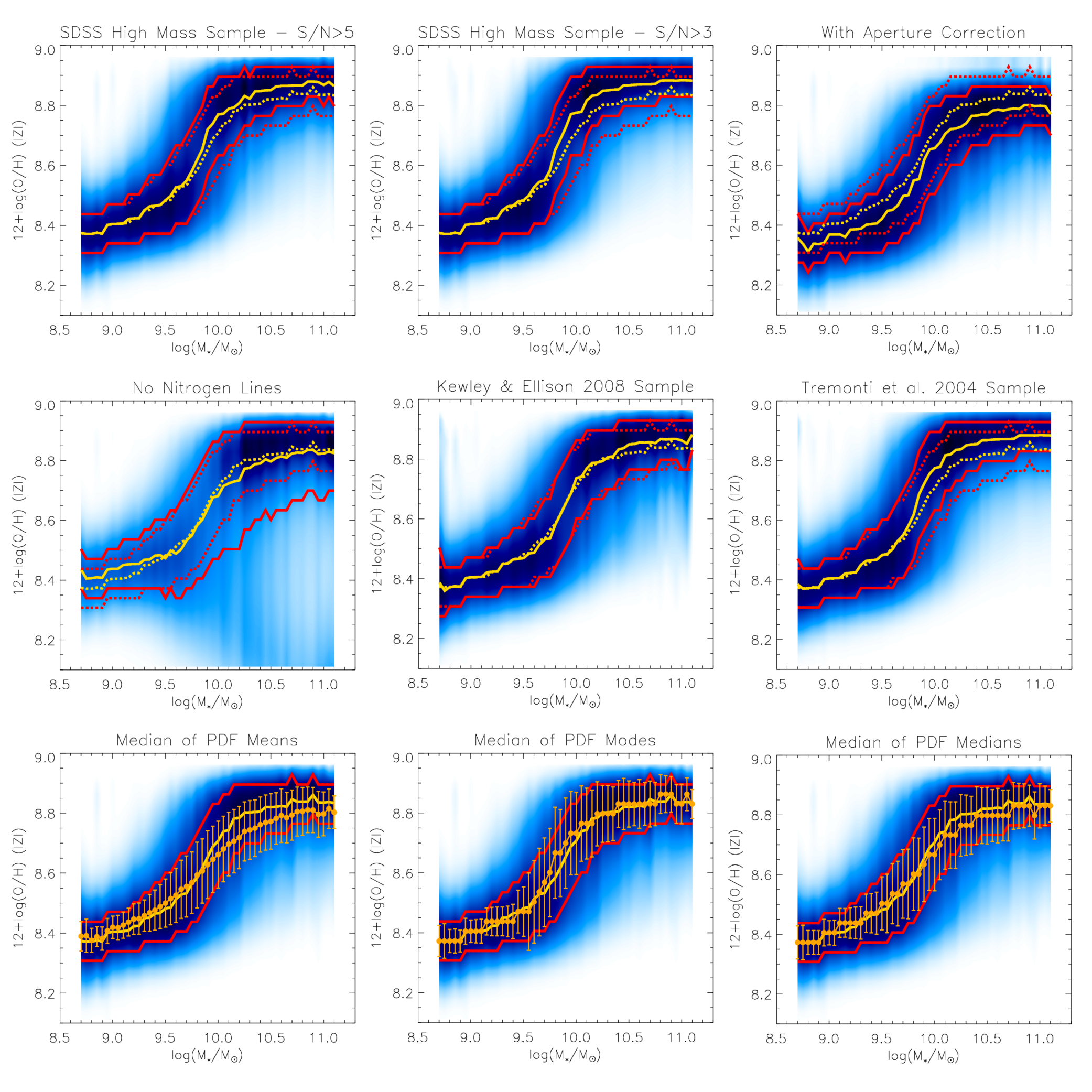}
\caption{Results from the tests described in Section \ref{sec:discussion-tests}. In the top two rows the dashed lines show the MZR of the SDSS high mass sample presented in Figure \ref{fig:mzr-sdss} for comparison. {\it Top Row:} The left and central panels show the MZR measured after changing the emission line $S/N$ threshold of the SDSS high mass sample ($S/N>5$ and $S/N>3$ respectively). The right panel shows the result of applying an aperture correction to the MZR following Equation 1 and assuming a flat surface brightness distribution. {\it Middle Row:} The left panel shows the resulting MZR after removing the [NII] doublet from the {\tt IZI} fits. The central and right panels show the MZR measured after mimicking the sample selections of \cite{kewley08} and \cite{tremonti04} respectively. {\it Bottom Row:} In these panels the SDSS high mass sample from Figure \ref{fig:mzr-sdss} is shown by the solid lines. Orange circles with error-bars show the MZR built by taking the median of the abundance of individual objects, measured from the individual galaxies metallicity PDFs using three different statistics (mean, mode, and median).}
\label{fig:mzr-tests}
\end{center}
\end{figure*}

The leftmost panel in the middle row of Figure \ref{fig:mzr-tests} shows the result of this test. For high mass galaxies, a secondary, low probability, low abundance peak in the PDF appears. This is because the [NII] doublet plays an important role in terms of removing the degeneracies associated with the double valued nature of the $R23$ diagnostics based on the oxygen lines (see Figure 1 in \citealt{blanc15} and Appendix A.1 in \citealt{kewley08}). Regardless of this expected behavior, the bulk of the probability remains associated with the correct high abundance peak of the PDF, and overall the observed MZR does not change significantly by removing the nitrogen lines from the fit. This implies that the observed transition in the MZR is not related to changes in the N/O abundance as a function of stellar mass.

\subsubsection{Applying {\tt IZI} to Literature Samples}
\label{sec:discussion-test-samples}

In order to provide a test to demonstrate that the characteristic shape of the MZR in this work does not arise from sample selection effects, we reproduce the sample selections of \cite{tremonti04} and \cite{kewley08}, and apply our method to measure the MZR on these samples. \cite{tremonti04} selected galaxies in the SDSS DR7 MPA/JHU catalog with $1\sigma$ stellar mass errors smaller than 0.2~dex, $S/N_{H\alpha, H\beta, [NII]}\ge 5$, and $0.005<z<0.25$. \cite{kewley08} selected objects with $S/N_{H\alpha, H\beta, [NII], [OII], [OIII], [SII]}\ge 8$ and $z<0.1$. Both studies rejected AGN contaminated systems using the BPT diagram as done in this work. 

The resulting MZRs measured with {\tt IZI} for these samples are shown in the central and rightmost panels in the middle row of Figure \ref{fig:mzr-tests}. Except from a small offset towards higher abundances in high mass systems for the \cite{tremonti04} sample, which is caused by the lower $S/N$ requirement as discussed above, the sample selection does not seem to affect the observed MZR. 

As a final test we applied all the literature diagnostics outlined in Appendix A.1 of \cite{kewley08} to our SDSS high mass sample. We find that the MZRs measured by taking the median abundances from these diagnostics in our sample, agree within the errors with those measured in \cite{kewley08} and shown in Figure \ref{fig:mzr-comp}. Given these and all previous tests we conclude that the characteristic transition mass scale in the MZR that we observe in this work is not introduced by sample selection effects. 

\subsubsection{Metallicity PDF Characterization}

Another possible source of systematic uncertainty is associated to the methodology used to construct the MZR by summing the metallicity PDFs of individual galaxies, and then taking the mean of this summed distribution. It could be thought that systematic variations in the shape of the PDF as a function of either stellar mass or metallicity could introduce biases in the mean of the summed PDF. To check if the shape of the MZR is significantly affected by our methodology, we compare our measurement to an MZR built using the classical approach of taking the median of single best-fit values for the abundance of individual galaxies in different mass bins. We adopt three different statistics (the PDF mean, median, and mode) to characterize the metallicity PDF of individual galaxies by a single number. Then we construct the MZR by taking the median of these statistics in narrow stellar mass bins (0.05~dex) and measure the 1$\sigma$ scatter for the abundance in each bin. These are compared with our adopted method to construct the PDF in the lower panels of Figure \ref{fig:mzr-tests}.

These three statistics suffer from different relative biases depending on the shape of the PDF. For example, while they agree with each other for any single peaked symmetric distribution, in the presence of asymmetries or multiple peaks in the PDF the values of the mean, mode, and median will deviate from each other. Therefore, the observed agreement with the shapes of the MZRs measured with these three methods, combined with the small scatter seen in these three statistics within each stellar mass bin, indicate that systematic variations in the shapes of the individual galaxy PDFs are not strong. This also shows that our method of summing the individual PDFs does not bias the shape of the MZR with respect to what is measured when assigning a single abundance value to each object.

\subsubsection{Impact of the SFR on the MZR}
\label{sec:discussion-tests-sfr}

Figure \ref{fig:m-sfr} shows that two of the dwarf galaxy samples used in this work (the DEEP2 and low mass SDSS samples) are biased towards high SFRs at fixed stellar mass with respect to the local $M^*-SFR$ sequence defined by the high mass SDSS sample, and its extrapolation towards low masses. On the other hand the LV dwarf galaxy sample closely follows this relation. This behavior is somewhat expected as the latter sample is local and largely volume limited (although not complete), while the former two samples suffer from stringent cuts in nebular emission line luminosity.

The SFR has been proposed as a secondary parameter in the MZR by several authors \citep[e.g.][]{mannucci10, lara-lopez10}. The proposed dependence implies that higher SFR galaxies have lower ISM metallicity at fixed stellar mass, and that a significant reduction in the scatter around the MZR is seen when this third dimension is taken into account in a so called "fundamental" mass-SFR-metallicity relation (FMR). Under this scenario, a bias in SFR introduced by emission line luminosity cuts could translate into a bias in the shape and normalization of the measured MZR \citep{juneau14}. 

On the other hand, a series of recent studies using different datasets and methodologies bring into question the existence of the FMR, or find marginally significant dependencies between metallicity and SFR at fixed stellar mass, much weaker than originally proposed. \cite{hughes13} use a volume and magnitude limited sample of 260 well characterized nearby late-type galaxies, and find no correlation between SFR and the residuals around the MZR. \cite{sanchez13} using integral field spectroscopic data from the CALIFA survey report no secondary correlation with either SFR nor sSFR in the global MZR nor the "local" MZR (i.e. the relation between stellar mass surface density and local metallicity on HII region scales). \cite{salim14} re-analyzed the SDSS sample used in \cite{mannucci10} to reassess the existance and shape of the FMR and the impact that sample selection and the use of different metallicity and SFR indicators have on it. They find a lack of correlation for massive ($>10^{10.5}$~M$_{\odot}$) star forming galaxies, and a significantly weaker anti-correlation between the relative sSFR at fixed stellar mass and metallicity for intermediate and low mass ($<10^{10.5}$~M$_{\odot}$) systems without extreme star formation activity. While they find intermediate and low mass galaxies with extremely enhanced relative sSFR ($>0.6$~dex) to show a strong anti-correlation between these quantities, they find that the overall scatter in the FMR does not greatly decrease from the scatter in the MZR. More recently \cite{barrera17} used integral field spectroscopy of more than 1700 galaxies in the SDSS-IV MaNGA survey \citep{bundy15} and found no strong secondary relation of the MZR with either SFR or sSFR.

\begin{figure}[t!]
\begin{center}
\epsscale{1.2}
\plotone{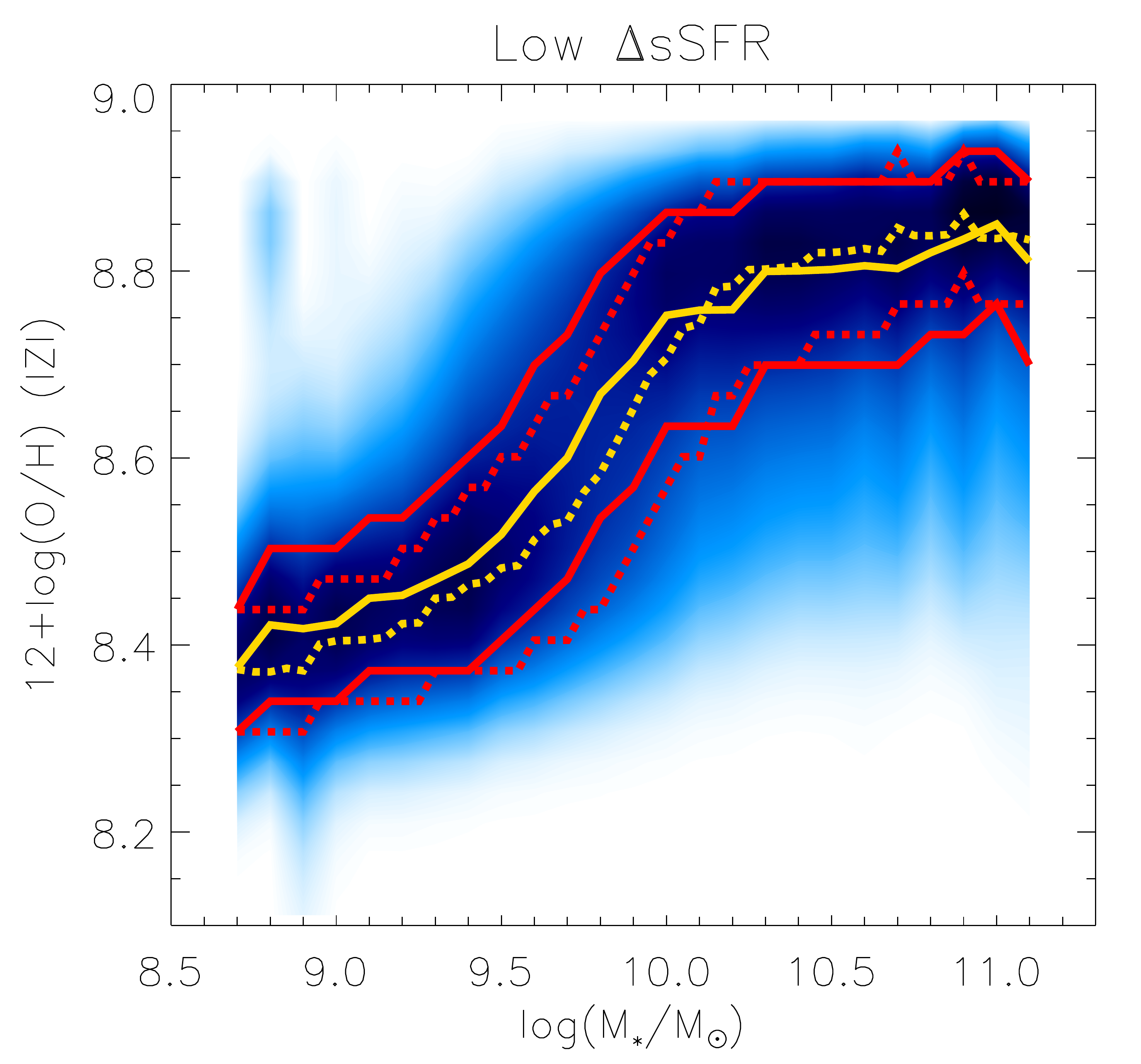}
\plotone{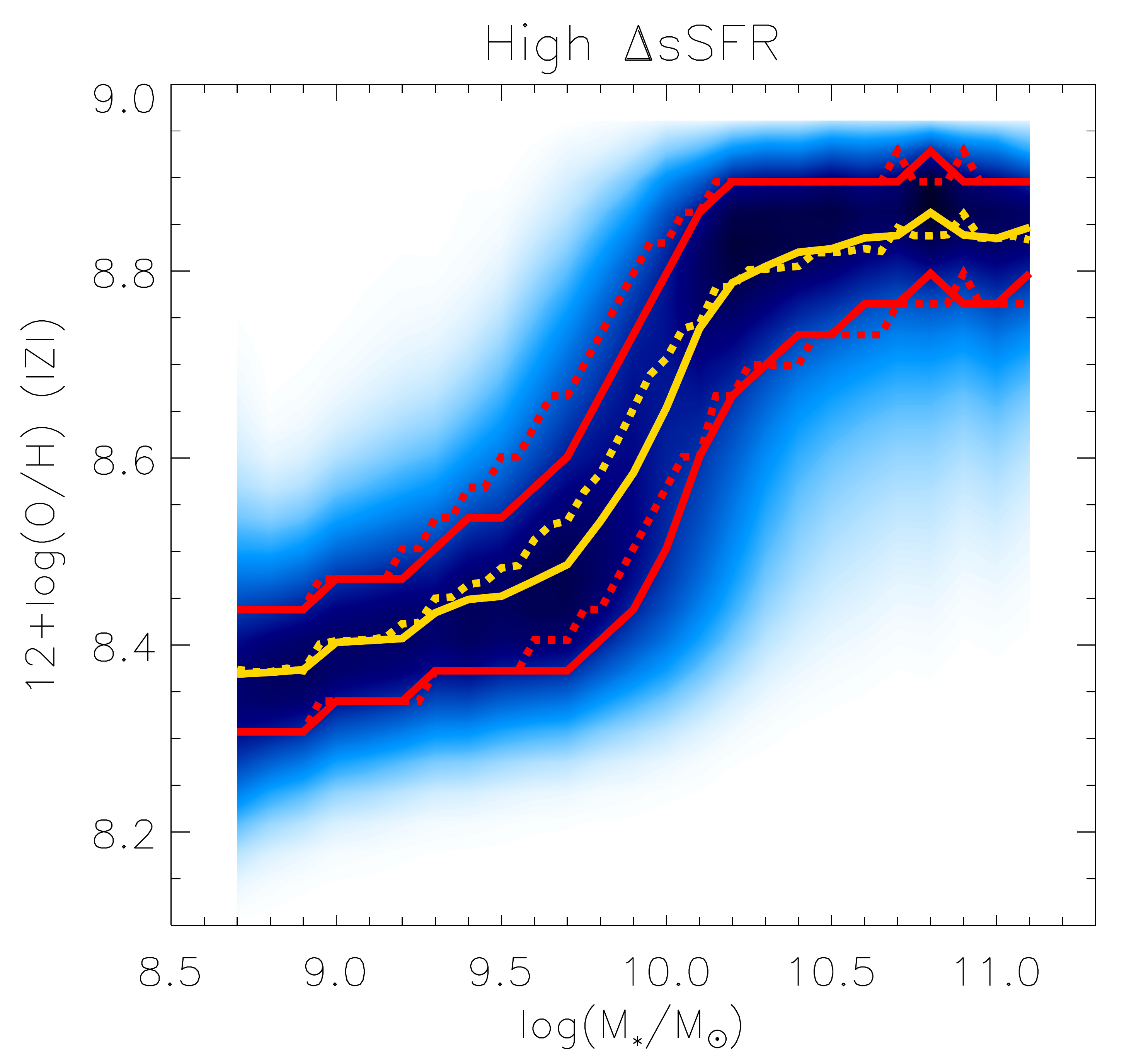}
\caption{SDSS high mass sample MZR after dividing it into systems with low (top) and high (bottom) star formation activity at fixed stellar mass. Colors are as in Figures \ref{fig:mzr-sdss} and \ref{fig:mzr-tests}, with solid lines showing the MZR of the low and high relative sSFR samples (i.e. galaxies below and above the dashed black line in Figure \ref{fig:m-sfr}), and dashed lines showing the MZR of the full sample.}
\label{fig:mzr-sfr}
\end{center}
\end{figure}

To quantify the impact that the SFR has on the shape and normalization of the MZR built under our methodology, we separate the SDSS high mass sample into two sub-samples with high and low relative sSFR and re-build the MZR following the same methodology used in Section \ref{sec:results-highmass}. The sub-samples are defined as galaxies having a sSFR either above or below the mean relation at any given stellar mass (dashed line in Figure \ref{fig:m-sfr}). While for the full sample MZR we adopted 0.05 dex bins in stellar mass, here we adopt 0.1 dex bins to account for the $\sim50$\% decrease in sample size. While a full study of the inter-dependencies between stellar mass, metallicity, and SFR is beyond the scope of this paper, this simple test highlights the robustness of our results with respect to SFR biases in sample selection.

Figure \ref{fig:mzr-sfr} presents the result of this test. No difference in the MZR is seen between the two sub-samples at $M^*>10^{10}$~M$_{\odot}$. At lower masses we do observe a deviation that follows the direction expected for a FMR, with high relative sSFR systems having lower abundance than their low relative sSFR counterparts. Over the $10^{9}$~M$_{\odot}>M^*>10^{10}$~M$_{\odot}$ range, in which the largest deviations are seen, the two samples show a median abundance offset of $0.05$~dex. In this same range the samples have a median offset in relative sSFR of $0.35$~dex. For such an offset, and at these masses and SFRs the \cite{mannucci10}, \cite{salim14}, and \cite{hughes13} formulations of the FMR predict abundance offsets of $\simeq 0.07$~dex, $\simeq 0.05$~dex, and $\simeq 0.02$~dex respectively. So our measurement provides a better match to the \cite{salim14} FMR, and is bracketed by the FMRs of \cite{mannucci10} and \cite{hughes13}. It is important to highlight that these offsets are well within the $1\sigma$ envelope of the MZR.  We cannot robustly discriminate between these different formulations of the FMR without a more detailed statistical analysis which is beyond the scope of this paper. Regardless of this, the magnitude of these offsets is not sufficient to significantly affect the observed change in slope of the MZR towards the low mass regime.

While for small relative sSFR offsets at intermediate masses our results are consistent with proposed formulations of the FMR, the agreement seen in Figure \ref{fig:mzr-dwarfs} between the MZR of the DEEP2, low mass SDSS, and LV dwarf samples at $M^*\simeq10^{8.0}$~M$_{\odot}$, despite their strong $0.5-1.0$~dex relative sSFR offsets, is largely inconsistent with this formulation. At these masses and SFRs the \cite{mannucci10} FMR predicts an abundance offset of $\sim0.2$~dex\footnote{The \cite{hughes13} and \cite{salim14} FMRs are not parameterized in a way that can be evaluated at these low masses.} while the MZR of these samples agree with each other to $\sim0.05$~dex. It is important to highlight that the \cite{mannucci10} FMR is not calibrated against data in this mass and SFR regime. The above prediction comes from an extrapolation of the relation fitted at higher mass and lower sSFR. 

In summary, considering the agreement seen in the slope of the low mass end of the MZR across different samples (despite their sSFR offsets), the fact that the observed break in the MZR appears regardless of us dividing the high mass SDSS sample in low and high relative sSFR subsamples, and that the LV dwarf MZR follows the same $M*-SFR$ relation as the high mass SDSS sample while showing a shallow slope in the low mass regime, we can confidently say that our result is robust against possible secondary dependencies of the MZR with the SFR.

\subsubsection{Impact of Ensemble Averaging and DIG on Integrated Metallicity Measurements}
\label{sec:discussion-test-dig}

In Section \ref{sec:results-lowmass} we find consistent results when comparing the MZR of LV dwarf galaxies (based on slit spectroscopy of individual HII regions) with the MZR of the high-mass SDSS sample (based on integrated central measurements as sampled by the SDSS fibers). The agreement is even better than shown in Figure \ref{fig:mzr-dwarfs} when considering the fiber aperture corrections we compute in Section \ref{sec:discussion-test-aperture} (top right panel in Figure \ref{fig:mzr-tests}). This indicates that at the stellar masses where the two relations meet, the process of integrating the emission over kpc scale areas in the centers of galaxies does not introduce strong biases on the measured metallicities.

Mixing luminosity weighted ensembles of HII regions with constant metallicity but with a dispersion in electron temperature has been shown to induce an underestimation of direct method oxygen abundances, while leaving SEL abundances mostly unchanged. This is a similar effect to the one introduced by electron temperature fluctuations in nebulae \citep{peimbert67}. The impact of such averaging on SEL fluxes and ratios is calculated to be small though, at a $<0.1$~dex level \citep{pilyugin12b, sanders17}.

DIG contamination on the other hand can significantly bias SEL ratios and metallicities derived from them if its contribution to the total integrated flux is high \citep{sanders17}. Early determinations of the DIG fraction in nearby spiral galaxies used narrow-band H$\alpha$ imaging to separate discrete sources from diffuse emission, finding high DIG fractions in the 40\%-60\% range \citep[e.g.][]{hoopes96, thilker02, oey07}. These high values are adopted in the recent work of \cite{sanders17}, who proposes that DIG contamination significantly affects the measurement of metallicities from SDSS spectra.

Recent results from intergral-field spectroscopy of nearby galaxies hint at the large DIG fractions measured from narrow-band studies being  largely overestimated, especially for the central kpc scale regions of spiral galaxies. As discussed below this implies that DIG contamination has a very small effect on the SEL abundances we measure with {\tt IZI} for the SDSS high-mass sample.

Narrow-band H$\alpha$ imaging typically suffers from low surface brightness sensitivity compared to modern IFU datasets. This can cause undetected low surface brightness HII regions to be erroneously classified as DIG. For example the \cite{oey07} SINGG data reaches a 1$\sigma$ line flux depth of $2\times10^{-17}$~ erg/s/cm$^2$/arcsec$^2$. In comparison, datasets like VENGA \citep[The VIRSU-P Exploration of Nearby Galaxies,][]{blanc13b} and PHANGS-MUSE \citep[Physics at High Angular Resolution in Nearby Galaxies,][]{kreckel16, kreckel18} typically reach depths of $5\times10^{-18}$~ erg/s/cm$^2$/arcsec$^2$ (i.e. a factor of 4 deeper in surface brightness). Furthermore, spectroscopic information in IFU datacubes allows for a better decomposition between DIG and HII regions than can be achieved using H$\alpha$ emission alone, due to the different nebular spectrum of these components. Finally, narrow-band H$\alpha$ filters include contamination from the [NII] doublet. This is typically corrected assuming a constant [NII]/H$\alpha$ ratio characteristic of HII regions. Since DIG shows enhanced [NII]/H$\alpha$ ratios with respect to HII regions, the H$\alpha$ flux of the diffuse component in narrow-band studies is most likely overestimated by up to $\sim80$\% \citep[see Figure 15 in][]{blanc09}.

\cite{blanc09} studied the contribution of diffuse ionized gas emission with enhanced [SII]/H$\alpha$ ratios in the central $4\times4$~kpc$^2$ of M51 using data from the VENGA survey \citep{blanc13b}. They find a DIG fraction in this region of only 11\%. This comes partly form the fact that DIG becomes more releveant towards the outer parts of galaxies where low surface brightness regions are more common, and partly from the fact that some low surface brightness regions that would have been below the detection threshold of narrow-band studies show HII-region like line ratios that reveal their star forming nature. Notice that the SDSS fiber diameter at the median redshift of the high-mass sample ($\langle z \rangle = 0.077$) is 4.4 kpc, similar to the area studied in \cite{blanc09}. Adopting DIG fractions measured across the full extent of galaxies as representative of their central regions will likely result in an overestimation of the DIG contribution to the total flux.

More recently \cite{kreckel16} used a mosaic of VLT-MUSE datacubes on NGC 628 (M 74), taken as part of the PHANGS survey, to study the differences between inter-arm and on-arm HII regions. They find that 64\% of all H-alpha flux in the inter-arm regions of NGC 628 can be resolved into discrete low luminosity HII regions in the deep MUSE data. Most of this flux would have been assigned to DIG in narrow-band studies like \cite{oey07}.

\cite{zhang17}, using data for 365 face-on galaxies in the SDSS-IV MaNGA survey, study how oxygen abundance diagnostics change as a function of H$\alpha$ surface brightness. They specifically tested {\tt IZI} and other seven SEL methods. Figure 28 in \cite{zhang17} shows that when using {\tt IZI}, the observed differences in $12+\log{(O/H)}$ from the lowest to the highest surface brightness regions at fixed galactocentric radius are typically at the $\pm 0.1$~dex level. Since the SDSS fibers sample a broad distribution of H$\alpha$ surface brightness within single galaxies, the expected biases in 12+log(O/H) due to DIG contamination will necessarily be $<0.1$~dex. In line with this, even when assuming a relatively high DIG contamination fraction, \cite{sanders17} finds typical biases in the SEL MZR for different methods that are $<0.1$~dex in most cases.

At MaNGA's spatial resolution ($\sim 1$~kpc) individual HII regions cannot be resolved. \cite{ho17} study the presence of azimuthal abundance variations in the spiral galaxy NGC 1365 using a spectral datacube with a spatial resolution of 150 pc from the TYPHOON project. They find a systematic trend with $\sim\pm 0.1$~dex variations in the oxygen abundance of discrete HII regions when going from inter-arm to on-arm environments. The trend goes in the same direction and has a similar magnitude as the trend observed in \cite{zhang17} for {\tt IZI}. This opens the possibility that some part of the variations seen in \cite{zhang17} could be due to azimuthal variations in the HII region oxygen abundances. This would imply an even lower impact of DIG contamination on integrated measurements of the oxygen abundance.

The above results are also in line with the direct test conducted by \cite{kaplan16} using the VENGA VIRUS-P datacube of NGC 2903, in which DIG regions are spectroscopically separated from HII region dominated areas. The metallicity gradient of the galaxy is then calculated using six different SEL diagnostics, both including and removing the DIG. The authors find that, for all the adopted methods, including DIG dominated regions only changes the normalization of the oxygen abundance gradient by at most $\pm 0.1$~dex, and its slope by $\pm 0.015$~dex~kpc$^{-1}$.

In summary, recent work based on IFU spectroscopy implies that the expected impact of DIG contamination on SEL oxygen abundance determinations in local star forming galaxies is small ($<0.1$~dex). Especially considering that the SDSS fibers typically sample the central few kiloparsecs of galaxies. Therefore we expect our results to be robust against the effects of HII region ensemble averaging and DIG contamination

\subsection{Systematic Offsets between {\tt IZI} and Direct Method Abundances}
\label{sec:discussion-test-abundances}

In the previous section we showed that the difference between the shape of the MZR in this work and previous studies that report a single power-law behavior does not arise from an exhaustive list of potential systematic sources of error related to sample selection, measurement biases, and adopted methodology. This implies that the origin of the discrepancy likely lies on systematic differences between the abundance diagnostics adopted in different studies.

The systematic differences between different SEL diagnostics and their impact on the shape of the MZR were thoroughly studied by \cite{kewley08} and have already been discussed in Section \ref{sec:results} (see Figures \ref{fig:mzr-comp} and \ref{fig:mzr-dwarfs-lit}). Here we study the differences between the oxygen abundances determined by {\tt IZI} and those determined from the direct method. This is of particular importance as previous measurements of the MZR in SDSS galaxies using the direct method find a single power-law behavior towards low masses \citep[e.g.][]{andrews13}.

\begin{figure}[t!]
\begin{center}
\epsscale{1.2}
\plotone{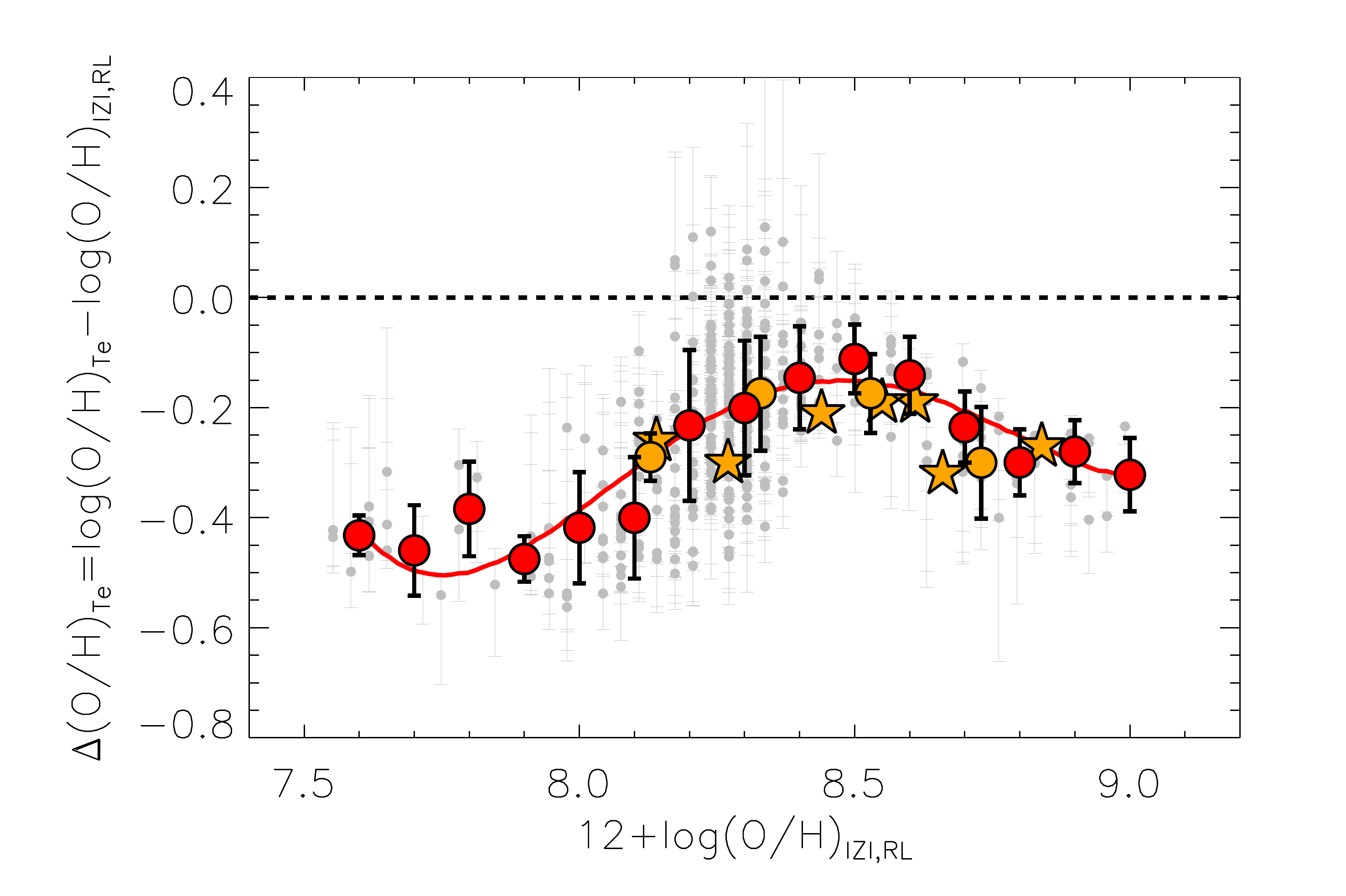}
\caption{Grey points with error-bars show the difference between the direct method and {\tt IZI} abundances for a sample of 414 HII regions in \cite{pilyugin12}. The median offsets and their 1$\sigma$ dispersion in 0.1~dex metallicity bin are shown by the red circles. Orange symbols show the offsets between direct method and RL abundances (the latter scaled by a constant 0.07~dex factor) from \cite{bresolin16} (stars), and \cite{toribiosancipriano17} (circles with error-bars).}
\label{fig:izi-te}
\end{center}
\end{figure}

At near-solar metallicities the direct method is known to produce oxygen abundances that are $\sim0.2$ dex lower than the results of other methods (RL and theoretically calibrated SEL diagnostics) if the effects of temperature inhomogeneities in the ionized gas are not corrected \citep[e.g.][]{aller54, peimbert67,peimbert69, garcia-rojas07, kewley08}. But a simple offset in abundance cannot explain the difference we see in the shape and slope of the MZR with respect to literature measurements that use the direct method. A metallicity dependent offset would be required.

To do a direct comparison between {\tt IZI} and direct method abundances down to the low metallicity regime we use a sample of 414 HII regions with direct abundance and SEL measurements compiled by \cite{pilyugin12}. For all objects in this sample we remeasure the oxygen abundance using {\tt IZI} to fit the [OII]$\lambda\lambda$3726,3729, H$\beta$, [OIII]$\lambda\lambda$4959,5007, [NII]$\lambda\lambda$6548,6584, and [SII]$\lambda\lambda$6717,6731 lines fluxes with the \cite{levesque10} photo-ionization models. This is the same setup used above to measure the MZR. Figure \ref{fig:izi-te} presents the comparison between the {\tt IZI} abundances and the direct method abundances measured by \cite{pilyugin12} for all regions in the sample (grey points with error-bars). Red circles in Figure \ref{fig:izi-te} show the median offset ($\Delta(O/H)_{Te}$) measured in 0.1~dex metallicity bins.

In the regime of solar to slightly sub-solar metallicities ($12+\log{(O/H)}\simeq 8.5$) we observe the offset of $\sim-0.2$~dex typically seen between direct method and theoretically calibrated SEL\footnote{Although notice that large scatter exists between different SEL diagnostics in the literature \citep{blanc15}} and RL abundances. Interestingly, the offset increases towards both lower and higher metallicities, reaching $\sim -0.3$~dex at $12+\log{(O/H)}\simeq 9.0$ and $\sim -0.4$~dex at $12+\log{(O/H)}<8.0$.

\begin{figure}[t!]
\begin{center}
\epsscale{1.2}
\plotone{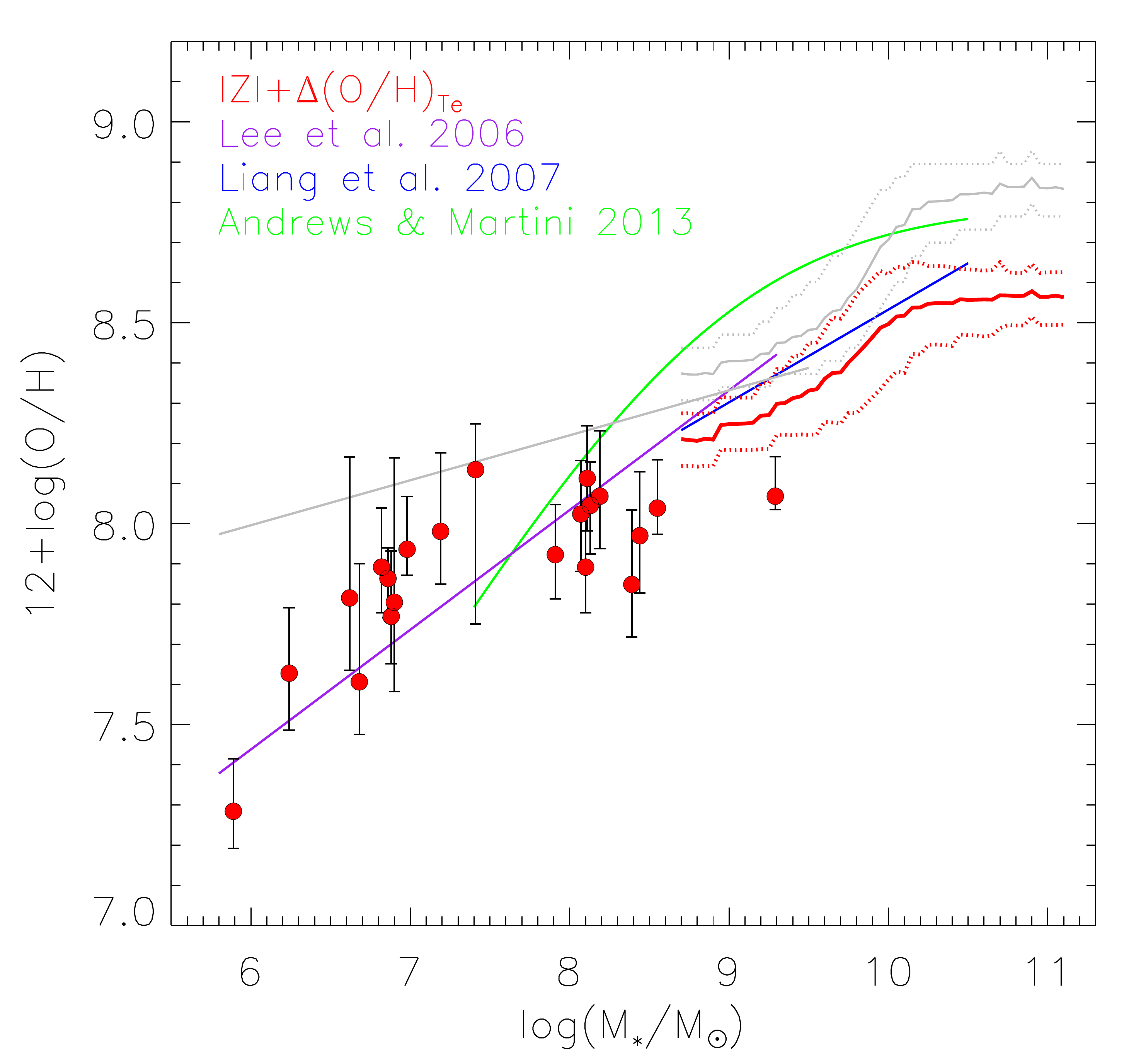}
\caption{MZR measured with {\tt IZI} scaled at each metallicity by the median offset between direct method and {\tt IZI} abundances ($\Delta(O/H)_{Te}$, red line in Figure \ref{fig:izi-te}). Also shown for comparison are the MZR measurements by \cite{lee06} for Local Volume dwarf galaxies, and by \cite{liang07} and \cite{andrews13} for stacks of SDSS galaxies in stellar mass bins. All based on direct method abundances. The non-scaled {\tt IZI} MZR is shown in gray.}
\label{fig:mzr-izi-te}
\end{center}
\end{figure}

This behavior between the two diagnostics can largely explain the discrepancies in the shape of the MZR measured using these two different methods. To show this is the case we fit a low order polynomial to the median offset as a function of metallicity (solid red line in Figure \ref{fig:izi-te}) and use it to scale the {\tt IZI} MZR shown in the left panel of Figure \ref{fig:mzr-dwarfs}. We compare the resulting MZR to the direct method based measurements of the MZR of \cite{lee06, liang07}, and \cite{andrews13} in Figure \ref{fig:mzr-izi-te}.

When applying this metallicity dependent offset the break in the MZR from the intermediate to the low mass regimes is softened significantly, making it consistent with a single power-law behavior towards low masses. This modified MZR agrees well within the error-bars with the direct method based measurements of \cite{lee06} and \cite{liang07}. The \cite{andrews13} MZR disagrees in shape and normalization with the other direct method MZRs, largely due to differences in the method used to estimate the low and high ionization zones electron temperatures, which translates into higher abundances. This difference reflects the current level of systematic uncertainty in direct method determinations of the MZR.

A puzzling inconsistency arises from the fact that the direct method produces a steeper low mass MZR than the $N2$ and $O3N2$ methods, which in Figure \ref{fig:mzr-dwarfs-lit} appear consistent with {\tt IZI}, since the latter diagnostics are calibrated against the former. The $N2$ and $O3N2$ calibrations in \cite{pettini04} are based on linear fits to a control sample of HII regions with direct abundance measurements. In the $12+\log{(O/H)_{T_e}} < 8.1$ range, which according to Figure \ref{fig:izi-te} corresponds to the $12+\log{(O/H)}_{IZI} < 8.3$ range below which the slope of the MZR flattens, it is evident from Figure 1 of \cite{pettini04} that the linear fit is shallower than the trend followed by the data points. A higher order cubic fit shown in the same figure better fits the data in this region and would have translated into a steeper calibration, and a steeper MZR in the low abundance regime. This could explain the discrepancy seen here between the $N2$ and direct method. Explaining the discrepancy with $O3N2$ is more difficult since \cite{pettini04} did not calibrate the relation below $12+\log{(O/H)_{T_e}}~\simeq8.2$ (corresponding to $12+\log{(O/H)}_{IZI} < 8.4$) because the scatter in the calibration sample increases significantly below this limit. Therefore we are using an extrapolation of the original calibration. Photo-ionization models show that $O3N2$ becomes less sensitive to metallicity in the low abundance regime (e.g. Figure 2 in \citealt{blanc15}), explaining the increasing scatter.

In light of the results presented in Figure \ref{fig:mzr-izi-te} and the tests outlined in previous sections, we can confidently conclude that the differences between the shape of the MZR presented here and classic results in the literature finding a single power-law behaviour towards low stellar masses come from a systematic difference in the methods adopted to measure nebular abundances.

\section{Discussion}\label{sec:discussion}
\label{discussion}

The presence of a break in the slope of the MZR at a characteristic transition mass scale has important implications for theories of galaxy evolution and chemical enrichment. Having established that this feature arises only when the MZR is measured using certain abundance diagnostics ({\tt IZI}, $N2$, and $O3N2$) and not others ($R23$ and the direct method), the obvious question that follow is: which of these methods are right?

In this section, we discuss the agreement between {\tt IZI} abundances and those measured directly from oxygen recombination lines, and show that these largely independent methods present similar systematic differences with respect to direct method abundances. We also compare the MZR we derive here with previous observational estimates in the literature and recent results from state of the art hydrodynamical simulations which do show the presence of an equivalent transition mass scale. The results of these comparisons support the reality of the features we observe in the MZR. Finally we discuss a possible physical origin for the MZR characteristic transition mass scale.

\subsection{Agreement between {\tt IZI} and Recombination Line Abundances}
\label{sec:discussion-RL}

The existance of potential systematic biases in direct method abundances has been recognized for decades \citep[e.g.][]{peimbert69, stasinska02, penaguerrero12}. Small scale temperature fluctuations and large scale temperature gradients in ionized nebulae, combined with the different temperature dependence of auroral and strong lines can induce a bias in the derived electron temperatures. This bias is such that direct method abundances are typically lower than the real values, unless a correction for these effects is taken into account. Such corrections are seldom applied though, and direct abundances like those in \cite{pilyugin12, andrews13, liang07} and \cite{lee06} do not include them.

Several authors have argued that this bias can be quantified by comparing direct method abundances with those derived from metal recombination lines \citep[RL, e.g.][and references therein]{toribiosancipriano16, toribiosancipriano17}. For the same regions RL abundances of oxygen typically appear between 0.15 and 0.45~dex higher than direct method abundances, and the logarithmic difference between the two is typically referred as the ''abundance scale discrepancy factor'' (ADF)\footnote{Except for rare exceptions in which $O^{+}$ recombination lines have been detected \citep{garcia-rojas06} the ADF is typically measured for $O^{++}/H$ and assumed to be the same as for the total $O/H$ abundance.}. For a recent compilation of ADF measurements see Figure 4 in \cite{toribiosancipriano17}. This study also finds a metallicity dependence of the ADF, with the smallest offsets ($\sim0.2$~dex) seen around $12+\log{(O/H)}_{RL}\simeq8.5$ and an increasing ADF towards lower and higher abundances, reaching values of $\sim0.4$~dex near $12+\log{(O/H)}_{RL}\simeq8.2$ and 8.8.

This behavior is almost identical to the one we see between {\tt IZI} and direct method abundances in Figure \ref{fig:izi-te}. \cite{blanc15} showed that {\tt IZI} abundances computed using the \cite{levesque10} models used here, agree fairly well with RL abundances, showing a median offset of 0.07~dex and a scatter of 0.1~dex. In Figure \ref{fig:izi-te} we overplot the RL vs. direct method ADF values calculated by \cite{toribiosancipriano17} after applying a 0.07~dex offset to bring their RL abundances into a common scale with {\tt IZI} (filled orange circles). We also plot the results of \cite{bresolin16} who report both RL and direct method abundances for six HII regions of different metallicities. These datapoints show a behavior that is consitent with the metallicity trends reported in \cite{toribiosancipriano17}.

This comparison shows that both RL abundances and the {\tt IZI} abundances used in this work show systematic deviations from direct method abundances that are not only similar in direction and normalization (to within a 0.07~dex factor), but also show a common dependence on metallicity. While this still leaves open the possibility that both {\tt IZI} and RL abundances are wrong, and direct method abundances are right, the fact that these two largely independent methods (RL and {\tt IZI}) agree is encouraging.

Insight on the difference between RL and direct method abundances can in principle be obtained from comparison to stellar oxygen abundances of young stars in the same star forming regions, or at similar galactocentric radii in the same galaxies. Unfortunately these studies currently yield contradictory and non-conclusive results.

In the nearby spiral NGC 300 \cite{bresolin09} find good agreement between HII region direct method oxygen abundances (with no correction for temperature fluctuations nor oxygen depletion on dust) and stellar oxygen abundances of B and A supergiants. \cite{simondiaz11} on the other hand compare direct method and RL abundance measurements of the Orion Nebula with stellar abundances in 13 B-type stars in the region. After correcting for the depletion of O on dust grains they find good agreement with RL abundances and not with the direct method. The later are lower by $\simeq0.1$~dex ($\simeq0.2$~dex before the dust depletion correction). In M31 \cite{zurita12} find that the direct method abundances of HII regions are typically $\sim0.3$~dex lower than the stellar abundances of BAF supergiants. \cite{garciarojas14} compared direct method oxygen abundances across the Cocoon Nebula (IC 5146), a nearby HII region in the Solar neighborhood, with the stellar oxygen abundance of its central ionizing B0.5 V star. They find the direct method to underpredict the stellar abundance by $\simeq0.2$~dex ($\simeq0.1$~dex after correcting for dust depletions).

The results of \cite{toribiosancipriano16, toribiosancipriano17} also show puzzling discrepancies in this respect. They find that RL oxygen abundances agree well with B supergiant abundances in NGC 300, but appear overestimated in M33, the LCM, and the SMC, where the direct method abundance show a better match. Finally, \cite{bresolin16} find that the photospheric oxygen abundance of A-type supergiants associated with HII regions in their sample agree better with the direct method ISM abundances than with RL abundances, with the RL abundances being increasingly overestimated towards lower metallicities. It is worth noticing that the spectral models they use to fit stellar spectra assume a solar abundance pattern. Considering that chemical evolution models and observations \citep[e.g.][]{yates13, nicholls17} show a decrease in O/Fe as a function Fe/H of about 0.2-0.3 dex over the range of metallicities explored in \cite{bresolin16}, and that the optical spectra used in that work is dominated by Fe absorption features, could in principle bring their stellar abundances in better agreement with RL abundances.

In summary, the comparison to stellar abundances has not provided conclusive evidence on whether the direct method is to be preferred over RL abundances or viceversa. As mentioned above, the agreement between {\tt IZI} and RL abundances highlighted in Figue \ref{fig:izi-te} is encouraging, and supports the idea that the break we observe in the MZR measured with {\tt IZI} is indeed a real feature. Notice that while RL abundances are not available at the lowest end of the metallicity range studied in this work, the range over which this agreement can be observed brackets the $12+\log{(O/H)}\simeq8.5$ region where the observed change in slope in the MZR happens.

\subsection{Comparison with Previous Observational Results}
\label{sec:discussion-literature}

The morphology of the MZR measured in this work is qualitatively different from that reported in several previous studies that find a single power-law behavior at low and intermediate stellar masses, followed by a plateau in metallicity towards the high mass end  \citep[e.g.][]{tremonti04, mannucci10,andrews13, zahid14a}. Particularly, the MZR slope of $0.14\pm0.08$ we observe in the low stellar mass regime is shallower than previously reported. 

For low mass galaxies \cite{lee06} measure a slope of $0.30\pm0.03$ using a literature compilation of direct method abundances. \cite{berg12} measures a consistent value of $0.29\pm0.03$ using the direct method on a similar sample with homogeneous spectroscopy. \cite{zahid12} find a slope of $0.27\pm0.02$ for a compilation of direct method abundances from \cite{lee06} and \cite{zhao10}, and a steeper slope of $0.47\pm0.01$ using the $R23$ diagnostic of \cite{kobulnicky04} on the low mass end of the MZR measured in a SDSS sample equivalent to our high mass sample. \cite{zahid12} do not fit for the slope of the MZR of their DEEP2 sample (used here), for which they use a variation of the $N2$ method of \cite{pettini04} that includes a correction for the N/O abundance. Inspection of their Figure 12 shows that this sample follows a shallower slope similar to the one we measure here with {\tt IZI} and the $N2$ methods, and deviates from the extrapolation to low masses of the SDSS MZR. All these discrepancies are consistent with the trends we see between different diagnostics in Figures \ref{fig:mzr-dwarfs-lit} and \ref{fig:mzr-izi-te}.

On the other hand, some studies in the literature do show the presence of a characteristic mass scale in the MZR. Studying the luminosity-metallicity relation (LZR) of local galaxies down to the dwarf regime, both \cite{salzer05} and \cite{rosenberg06} report a flattening of the LZR for galaxies fainter than $M_B=-18$. Assuming a $B$-band mass-to-light ratio of $\Upsilon_{B}=1.0$ \citep[typical of a blue stellar population,][]{bell03} this transition corresponds roughly to a stellar mass of $\simeq 10^{9.4}$~M$_{\odot}$. This is close to the point where we see a flattening in the MZR. Given the small sizes and selection biases affecting these samples, the evidence for the break is tentative at best. More recent work by \cite{zhao10} does not find a break in the LZR of dwarf galaxies, but their sample includes very few galaxies brighter than $M_B=-18$ making it hard to trace a change in the slope above this limit.

\cite{sweet14} use the [NII]/[SII] vs. [OIII]/[SII] metallicity diagnostic of \cite{dopita13}, to study the R-band LZR of SDSS star forming galaxies and dwarf galaxies in the CHOIRS survey \citep{sweet13} and the literature. They propose the presence of two distinct populations in the SDSS LZR which follow relations with different slopes. Towards faint magnitudes dwarf galaxies with no evidence of having a tidal origin follow a shallower
relation \citep[see Figure 6 in ][]{sweet14}. Their bright and intermediate SDSS populations, following flat and steep LZRs respectively, and their dwarf galaxy population following a shallow relation are fully consistent with the three distinct regimes that we observe in the MZR.

Recently, \cite{kashino16} measured the SDSS MZR using the metallicity diagnostic proposed by \cite{dopita16}, which is based on the use of the [NII]$\lambda$6584/[SII]$\lambda$6717 and [NII]$\lambda$6584/H$\alpha$ ratios. They find a behavior consistent with the one reported here, with a
shallow slope for the MZR at low mass, followed by a steeper relation, and then a flattening at higher masses. In their case the steepening begins at a slightly lower mass than observed here ($\simeq10^9$ M$_{\odot}$ instead of $\simeq10^{9.5}$ M$_{\odot}$). The authors however, dismiss the
significance of the flattening towards low masses and interpret it as an artifact of the metallicity diagnostic. They argue that in the primary nitrogen production regime expected for low mass galaxies the [NII]/[SII] ratio will tend to a constant value,  artificially flattening the relation. 

We do not agree with the above interpretation for several reasons. First, the \cite{dopita16} calibration is based on photo-ionization models that include a transition between a primary and a secondary nitrogen production regime. This is introduced in the models as an empirically based average relation between the N/O and O/H abundances. Second, the method does not rely on the use of the [NII]/[SII] ratio alone, but also on the [NII]/H$\alpha$ ratio which will maintain a dependence on metallicity even in the primary nitrogen production regime. Therefore, as shown in Figure 3 of \cite{dopita16} a good correlation between the adopted diagnostic and the oxygen abundance should hold down to very low metallicities ($12+\log{(\rm{O/H})}\sim7.7$), significantly lower than the metallicities \cite{kashino16} measures at the low mass end of the SDSS sample  ($12+\log{(\rm{O/H})}\sim8.2$). Furthermore, as discussed in Section \ref{sec:discussion-test-n/o} when running our abundance estimation for the SDSS sample without using any nitrogen lines the break is still present (midle left panel of Figure \ref{fig:mzr-tests}.

An independent line of evidence for the existence of a break in the MZR comes from the stellar mass-metallicity relation of galaxies. In particular the results of \cite{gallazzi05} and \cite{kirby13} measuring the iron abundance of SDSS galaxies and stars in dwarf spheroidal and irregular systems in the Local Group, show a similar behavior to the one we observe for the gas-phase MZR. Inspection of Figure 9 in \cite{kirby13} shows that while the dwarf galaxy stellar MZR joins smoothly with the low mass end of the SDSS MZR, a steepening of the slope of the stellar MZR is observed around $10^{10}$ M$_{\odot}$. The stellar and gas-phase relations are not necessarily expected to trace each other. The former depends more critically on the star formation history of the galaxy, while the latter is more dependent on the recent balance between gas accretion, star formation, and outflows. Regardless, observing the same transition mass scale in both the stellar phase and gas-phase MZR strengthens the idea that it corresponds to a real feature. While the above comparison is compelling, the \cite{kirby13} result should be contrasted with the recent work by \cite{kudritsky16} and \cite{zahid17}, who find a continuous slope in the stellar MZR across the $10^7-10^{10}$~M$_{\odot}$ range.

\subsection{Comparison with Numerical Simulations}

Theoretical support for the existance of a transition mass scale in the MZR can be found in both cosmological and zoomed-in hydrodynamical simulations of galaxy formation. \cite{dave11b} run N-body + smoothed particle hydrodynamics (SPH) simulations with a box size of 48 Mpc and sufficient resolution to study the gas-phase MZR down to a stellar mass of $1.1 \times 10^9$ M$_{\odot}$ (i.e. 64 stellar particles). The authors use four different feedback prescriptions. One with no galactic winds, two prescriptions with constant velocity winds, and a momentum driven wind prescription with a mass loading factor inversely proportional to the galaxy velocity dispersion. Inspection of Figure 9 in \cite{dave11b} shows that except for the unrealistic model with no winds, all the other prescriptions predict a MZR that follows a qualitatively similar behavior to the one we find
using {\tt IZI}. That is, a flat saturated relation at high masses, followed by a steep relation at intermediate masses and then a flattening below $10^{9.5}$ M$_{\odot}$. In the simulations this trend is independent of environment. While this is encouraging, the flattening at low stellar masses happens very close to the resolution limit of the simulation. So it is hard to consider this a robust result.

\cite{schaye15} present the results of the Virgo Consortium Evolution and Assembly of GaLaxies and their Environments
(EAGLE) project. These are a suite of N-body+SPH hydrodynamical cosmological simulations of galaxy and supermassive black hole formation that implement different sets of feedback prescriptions in boxes of different sizes and with different spatial resolutions. Their ``reference'' and ``high resolution'' runs reach one and two orders of magnitude lower stellar masses than the \cite{dave11b} simulations respectively. 

\begin{figure}[t!]
\begin{center}
\epsscale{1.25}
\plotone{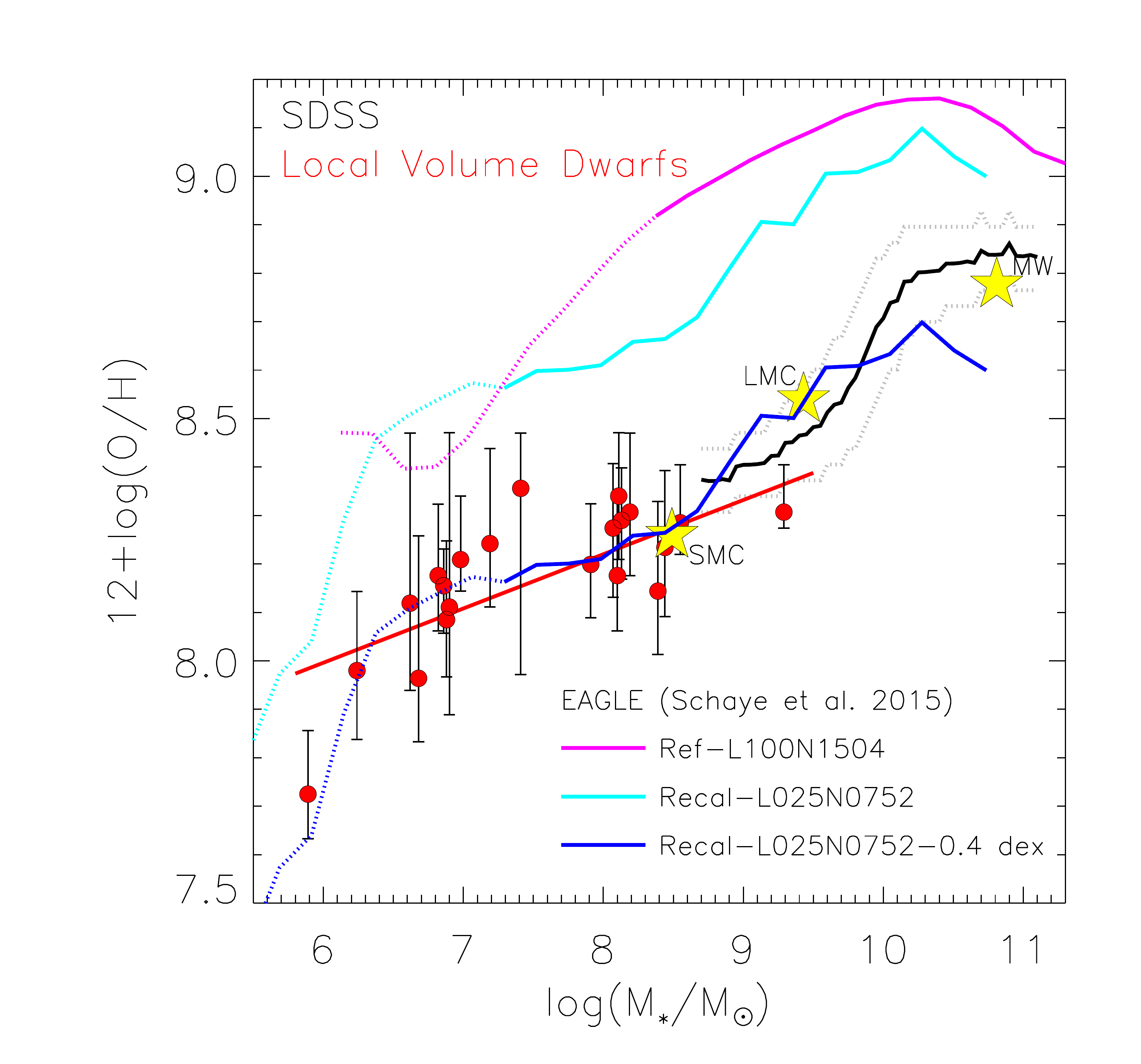}
\caption{Comparison between the MZR measured with {\tt IZI} for the SDSS high mass and the Local Volume dwarf samples, and the results from the EAGLE simulation \citep{schaye15}. The MZR for the lower resolution {\it Ref-L100N1504} run is shown in magenta, and that for the higher resolution {\it Recal-L025N0752} run, which is re-calibrated to match the $z=0$ GSMF is shown in cyan. The blue line shows this later run offset by -0.4~dex, to match the abundance scale of the {\tt IZI} measurements. The curves are dotted below the resolution limit of the simulations. The positions of the MW, LMC, and SMC galaxies in the MZR from high quality abundance determinations are marked by the yellow stars.}
\label{fig:mzr-izi-eagle}
\end{center}
\end{figure}

In Figure \ref{fig:mzr-izi-eagle} we compare our measured MZR to the results of two EAGLE runs presented in \cite{schaye15}. The \mbox{{\it Ref-L100N1504}} run is the reference run for the EAGLE project, with a box size of $100$~Mpc, $1504^3$ dark matter particles, and a baryonic particle mass of $1.81\times10^6$~M$_{\odot}$. The \mbox{{\it Recal-L025N0752}} is a high resolution run with recalibrated sub-grid stellar and AGN feedback parameters, tuned to improve the match to the $z=0$ galaxy stellar mass function (GSMF). This run has a box size of $25$~Mpc, $752^3$ dark matter particles, and a baryonic particle mass of $2.26\times10^5$~M$_{\odot}$. The measured $z=0$ gas-phase MZRs in these two simulations are shown by the magenta and cyan curves respectively. Curves are dotted below the mass limit of 100 baryonic particles per galaxy. This corresponds to $10^{8.26}$~M$_{\odot}$ and $10^{7.36}$~M$_{\odot}$ for the low and high resolution runs respectively.

Figure \ref{fig:mzr-izi-eagle} shows that both EAGLE runs predict abundances that are higher than those measured by {\tt IZI}, with the high resolution recalibrated run showing metallicities that are lower than the reference run as discussed in \cite{schaye15}. Interestingly, while the reference run does not show the presence of a break in the MZR, this feature appears in the high resolution recalibrated run, although at a mass scale $\sim0.7$~dex lower than observed in our measurement. Applying an offset of $-0.4$~dex to the {\it Recal-L025N0752} MZR brings it to agreement (within the errors) with the MZR measured with {\tt IZI}, except at the very high mass end where sampling effects due to the small box size start taking place \citep{schaye15}. Such an offset in the absolute abundance scale can be easily accommodated within the level of systematic uncertainty on the metallicities predicted by numerical simulations \citep{ly16, guo16b}. It is important to notice that the change in the morphology of the MZR in EAGLE is not only driven by the increase in resolution but it is also significantly driven by the fine-tuning of the feedback parameters as discussed in \cite{schaye15}.

While some of the \cite{dave11b} and \cite{schaye15} simulations hint to the presence of a transition mass scale in the MZR, it is important to stress that this is not a common feature of all cosmological hydrodynamical simulations in the literature. In fact, different simulations, run with different codes, methods, box-sizes, resolutions, and feedback prescriptions, present large discrepancies between each other regarding the shape and scatter of the gas-phase MZR \citep{ly16, guo16b, ma16}. This lack of consistency implies that the predictions of cosmological simulations regarding the chemical enrichment of star forming galaxies should be interpreted with caution, and stresses the importance of better observational determinations of the MZR, which is the goal of this work.

\begin{figure}[t!]
\begin{center}
\epsscale{1.25}
\plotone{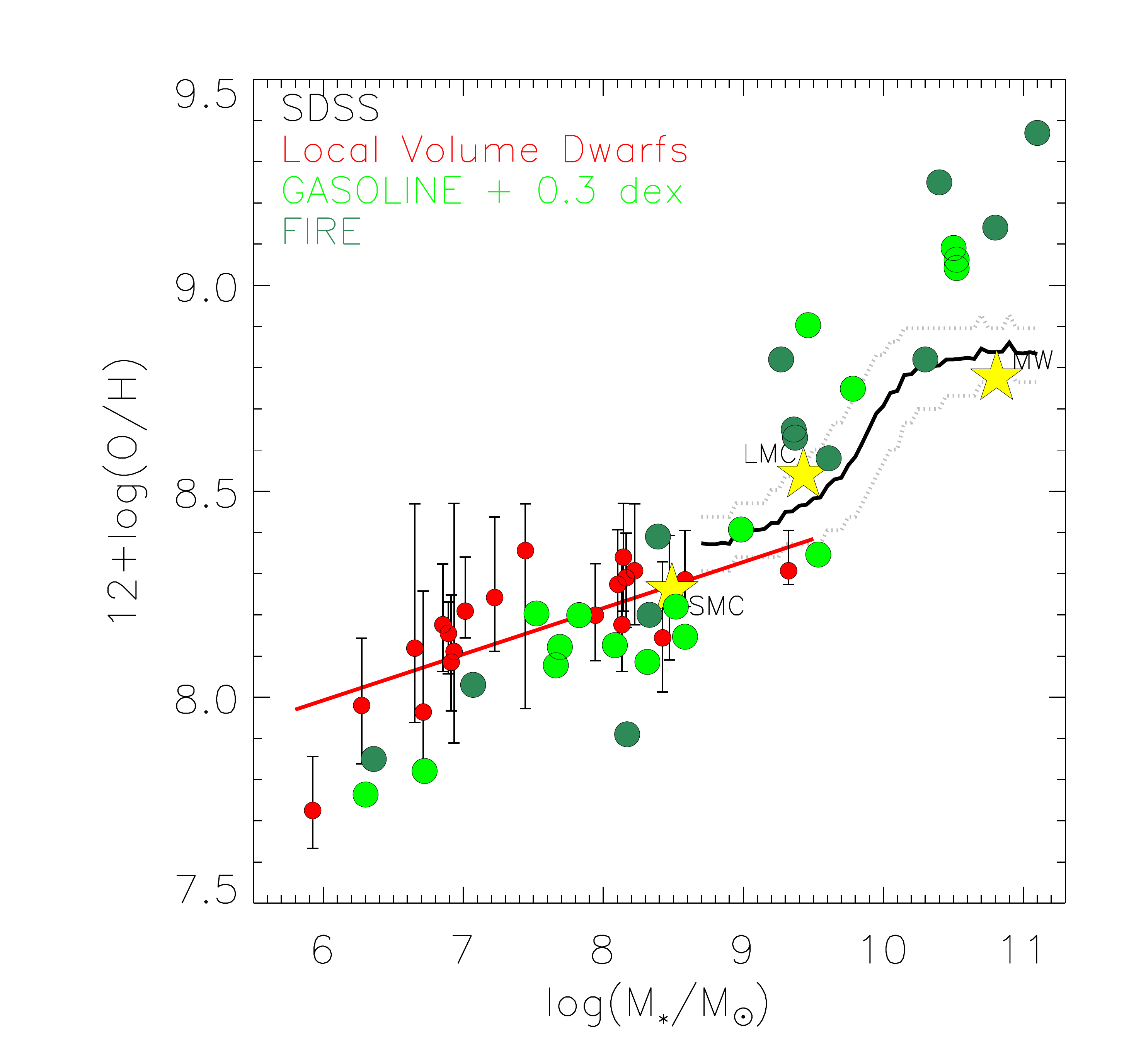}
\caption{Same as Figure \ref{fig:mzr-izi-eagle} but comparing to the results of the high resolution zoom-in hydrodynamical simulations of \cite{ma16} computed as part of the FIRE project (dark green circles), and \cite{christensen16, christensen18} which use the code GASOLINE (light green circles).}
\label{fig:mzr-izi-zoomin}
\end{center}
\end{figure}

Nevertheless the results of \cite{dave11b} and \cite{schaye15} show that it is possible for cosmological simulations to give raise to an MZR that has the features we observe in this work. Furthermore, the fact that the break in the MZR appears in the EAGLE run with improved resolution and feedback prescriptions, that allow the simulation to reproduce a range of observables \citep{schaye15}, is encouraging.

Besides cosmological simulations, which suffer from the effects of low spatial and mass resolution, it is interesting to see if the break in the MZR can be reproduced by zoom-in high resolution galaxy simulations. These type of simulations can typically be run on a much smaller number of galaxies but have the advantage of resolving the multi-phase ISM in more realistic ways that allow for better motivated star formation and feedback prescriptions to be implemented. The latter should translate into better predictions regarding the ISM chemical enrichment than can be achieved in cosmological simulations. Figure \ref{fig:mzr-izi-zoomin} compares our results to those from two recent studies that use zoom-in hydrodynamical simulations of galaxies to explore the MZR. Light green circles in Figure \ref{fig:mzr-izi-zoomin} show the simulations of \cite{christensen16, christensen18} and dark green circles show those of \cite{ma16}

The \cite{christensen16, christensen18} results come from N-body+SPH zoom-in simulations of 20 galaxies taken from a cosmological context using the code GASOLINE \cite{wadsley04}. The \cite{ma16} simulations also have realistic cosmological initial conditions and were run as part of the FIRE (Feedback In Realistic Environments) project. They include snapshots of 10 galaxies simulated with the P-SPH code GIZMO \cite{hopkins14}. We refer the reader to the original references for details regarding the different simulation codes, parameters, and prescriptions for star formation and feedback adopted in different runs. Here we only highlight that these simulations reach 1 to 3 orders of magnitude better resolution than the high resolution run of the EAGLE simulation discussed above (i.e. baryonic particle masses of $\sim10^2-10^4$~M$_{\odot}$ versus $\sim10^5$~M$_{\odot}$).

Both the \cite{ma16} and \cite{christensen18} simulations show hints for a change of slope in the MZR at $\sim10^9$~M$_{\odot}$. The absolute oxygen abundance level at $z=0$ in the \cite{christensen18} simulations is $\sim0.3$~dex lower at fixed stellar mass than in the \cite{ma16} simulations. The latter agree better with our measurements, so we scale up the metallicities in the former simulations by this amount to bring them to a common scale\footnote{This level of systematic discrepancy between simulations is not unexpected given the differences in the algorithms and prescriptions adopted to produce, eject, and diffuse metals in the ISM. These are linked to choices regarding SN rates, stellar mass loss rates, nucleosynthetic yields, the efficiency of feedback, etc.}. When brought to a common scale and put together in Figure \ref{fig:mzr-izi-zoomin} the change of slope in the MZR becomes evident. A power-law fit to the combination of both samples yields MZR slopes\footnote{No errors are provided for these slopes as no errors are reported for the simulations.} of $\alpha=0.23$ for $M_* \le 10^{9}$~M$_{\odot}$ and $\alpha=0.38$ for $M_* > 10^{9}$~M$_{\odot}$. While in the zoom-in simulations the break appears at lower masses than in our measurements ($\sim10^9$~M$_{\odot}$ versus $\sim10^{9.5}$~M$_{\odot}$), the behavior is qualitatively consistent with the one reported here and the slopes agree with the slopes we report for the low and intermediate mass regimes. The simulated MZR does not reproduce the flattening of the observed MZR at high stellar masses, which could be related to the lack of AGN feedback in the simulations \citep{ma16}.

While zoom-in simulations should be in a much better position to produce reliable predictions regarding the MZR than cosmological simulations are, mainly due to the fact that they resolve the structure and hydrodynamics of the ISM in significantly more detail, the fact that we see large normalization offsets between them highlights how sensitive their results are to the adopted feedback prescriptions and methodologies. As mentioned above, it important to interpret the results of simulations with caution. 
The level of systematic uncertainty in numerical simulations of galaxy chemical enrichment is currently too large to confidently use them as discriminators to asses the validity of different observational approaches. Nevertheless, it is encouraging to see that several state of the art high resolution simulations can indeed reproduce a feature similar to the one we observe in the MZR with {\tt IZI}.

\subsection{The Physical Origin of the Characteristic Mass Scale in the MZR}

The gas-phase MZR is the product of a complex interaction between several coupled physical processes. The main ones being: the accretion of gas from the IGM followed by star formation, the production of metals via nucleosynthesis, the injection of metals to the ISM via SN explosions and mass loss events during early (stellar winds from massive stars) and late (envelope shedding in post-AGB stars) stages of stellar evolution, the mixing of enriched gas in the turbulent ISM, the removal and ejection of metals towards the CGM and beyond the halo via feedback processes (i.e. fountains and outflows), the mixing of enriched and pristine gas in the CGM, and the re-accretion of recycled enriched gas. The efficiency with which these processes proceed in a galaxy ultimately determine the metal content of its ISM. Therefore, the existence of a characteristic mass scale at which a change of slope is seen in the MZR is likely associated with a stellar mass dependent transition in the efficiency of one or more of these processes.

The parameters affecting metal production and injection processes (e.g. the IMF, nucleosynthetic yields, SN rates, mass loss rates, etc.) are directly linked to stellar evolution processes which, beyond dependencies with metallicity itself and more subtle dependencies on the star formation history (e.g. alpha enhancements), are unlikely to show a sharp transition for galaxies at these stellar masses. IMF variations in particular have been studied as possible cause for giving rise to the shape of the MZR by \cite{spitoni10} who finds the need for an IMF that gets shallower in high mass galaxies, contrary to recent claims \citep[e.g.][]{conroy12, cappellari12}. Similarly, the mixing of newly injected metals in a turbulent ISM proceeds on very small scales ($\sim 10-100$~pc) and on short timescales \citep[$\sim10^8$~yr,][]{krumholz18}. It is therefore unlikely for the efficiency of these local scale processes to be behind the observed stellar mass dependent transition.

Larger scale effects like gas accretion and the efficiency which which gas can cool and settle into the star forming ISM, as well as the driving of galactic winds and the efficiency with which feedback can remove metals from the ISM, either ejecting them from the halo or mixing them with CGM gas for posterior re-accretion, are more likely to be linked to the observed transition mass scale in the MZR. All these processes are largely coupled with each other and it is hard to model their complex interactions separately. This makes hydrodynamical simulations a good tool to obtain qualitative and quantitative insight regarding the behavior of the large scale metal cycle in galaxies.

Both \cite{ma16} and \cite{christensen18} used their zoom-in galaxy simulations to study the ability of galaxies to retain or lose the metals they produce during their lives. Figure 8 of \cite{ma16} shows the fraction of produced metals that are retained within the virial radius of their simulated galaxies at $z=0$ as a function of their stellar mass. For low mass systems there is no significant trend in this quantity as a function of stellar mass, and a large scatter is observed with metal retention fractions between 0.3 and 0.8. Interestingly, for stellar masses above $\sim10^{10}$~M$_{\odot}$ this fraction tends towards unity indicating an increase in the ability of galaxies to retain their metals. The small number statistics provided by the less than a dozen simulated galaxies in \cite{ma16} prevent us from robustly stating that a sharp transition is in place around this stellar mass, but their results are certainly consistent with one and hint to its presence.

Similarly, \cite{christensen18} used their simulations to study the stellar mass dependence of the opposite quantity: the fraction of produced metals that are expelled beyond the virial radius of galaxies. They find that this fraction is fairly flat at stellar masses below $\sim10^{10.5}$~M$_{\odot}$. Above this threshold a clear trend of decreasing expelled fraction towards higher stellar masses appears. This is qualitatively consistent with the results of \cite{ma16}, and with a scenario in which galaxies at stellar masses around $10^{9.5-10.5}$ transition into a regime in which the interaction between star formation, feedback, metal ejection, mixing and recycling allows them to retain a much larger fraction of the metals they produce than can be retained in lower mass galaxies.

In this scenario, what actually causes a transition in the ability of galaxies to retain their metals remains an open question. An interesting possibility is associated with an observed transition in the morphology of the cold ISM of star forming galaxies and the way in which feedback dynamically couples to gas along different lines of sight. \cite{hayward17} propose an analytic model for how momentum deposition from stellar feedback in a turbulent ISM drives outflows and regulates star formation. In their model the ISM density presents a log-normal distribution (as expected in a turbulent medium) that allows for low surface density patches of gas to be accelerated beyond the escape velocity of the disk, even if the mean gas surface density is in the sub-Eddington limit. These low surface density patches can be efficiently ejected from the ISM in
the form of outflows while momentum injection into higher density parcels of gas drives turbulence which provides pressure support against gravity and regulates star formation \citep[see also][]{thompson16}. The main conclusion of \cite{hayward17} is that in galaxies with stellar masses above
$\sim10^{10}$ M$_{\odot}$ (which have low gas fractions of $f_g\lesssim 0.3$) outflows should be largely suppressed and the mass loading factor should drop sharply. This is because in a self-regulated quasi-stable turbulent medium the width of the density PDF becomes narrower as the gas fraction decreases, leaving a negligible fraction of the ISM in the super-Eddington regime once this threshold is reached.

As stated in \cite{hayward17} this transition in the efficiency with which momentum from star formation feedback couples to the ISM and drives winds is associated with a mass scale above which galaxies develop dynamically cold disks and the ISM becomes ``smoother'' (see also \citealt{muratov15}). Observational evidence for such a transition in the spatial structure of the cold ISM of disk galaxies is found by \cite{dalcanton04} who studied the distribution of dust in a sample of edge-on disk galaxies. The authors find that galaxies with rotation velocities $V_c\ge 120$ km s$^{-1}$ typically show well defined thin dust lanes in their mid-planes which reflect the fact that the cold ISM is distributed in a settled thin disk. On the other hand thin dust lanes are not present in $V_c< 120$ km s$^{-1}$ edge-on galaxies in their sample, which show a thick and clumpy distribution of dust (as thick as the stellar component). \cite{dalcanton04} interprets this sharp transition in the spatial structure of the cold ISM in terms of a
transition in gas disk stability. With the $V_c=120$ km s$^{-1}$ limit corresponding to the typical mass scale above which star forming
galaxies have the bulk of their disks in a gravitationally unstable regime. The stellar mass Tully-Fisher relation of local star forming galaxies \citep{mcgaugh05} implies that $V_c=120$ km s$^{-1}$ corresponds to a stellar mass of $\simeq 10^{10}$ M$_{\odot}$, in rough agreement with our observed transition mass scale in the MZR and the predicted transition mass scale in \cite{hayward17}.

In summary, it is likely that the characteristic mass scale we observe in the MZR at $\sim10^{9.5}$~M$_{\odot}$ corresponds to a point during galaxy growth above which feedback is no longer effective at ejecting metals from the ISM via outflows, and removing part of those metals from the halos of galaxies. Such drop in efficiency can be associated with a transition in the structure and stability of cold gas in star forming galaxies. With the ISM of galaxies above this mass threshold settling into dynamically cold and thin unstable disks, and the fraction of low column density super-Eddington lines of sight, easily accelerated and entrained into outflows, decreasing accordingly. The associated increase in the ability of galaxies to retain the metals they produce would translate into a process of rapid ISM enrichment, characterized by a steep MZR in this regime.

\section{Summary and Conclusions}\label{sec:summary}

In this work we measured the gas-phase MZR of local star forming galaxies across five orders of magnitude in stellar mass ($10^6$~M$_{\odot}<M_*<10^{11}$~M$_{\odot}$). We used {\tt IZI} to estimate the nebular oxygen abundance of galaxies selected from four complementary samples covering this large dynamic range in mass. Star forming galaxies are selected from SDSS DR7 and three samples of low mass systems from \cite{lee06, izotov06}, and \cite{zahid12}. {\tt IZI} uses Bayesian inference and photo-ionization models to derive abundances in ionized nebulae. The method and adopted photo-ionization models are presented and validated in \cite{blanc15}.
From this exercise we draw the following main conclusions:

\begin{itemize}

    \item The gas-phase MZR shows a complex morphology and behaves differently over three distinctive regimes in stellar mass. In the low mass regime ($M_{*}<10^{9.5}$~M$_{\odot}$) the MZR has a shallow power-law slope $\alpha=0.14\pm0.08$, which steepens sharply to $\alpha=0.37\pm0.08$ in the $10^{9.5}$~M$_{\odot}<M_*<10^{10.5}$~M$_{\odot}$ range, flattening down to $\alpha=0.03\pm0.11$ in the high mass regime ($M_*> 10^{10.5}$~M$_{\odot}$). While the flattening at high masses has been commonly observed and reported in the past, the transition between the low and intermediate mass regimes signals the existence of a previously unreported characteristic transition mass scale at $\sim 10^{9.5}$~M$_{\odot}$, above which, the level of chemical enrichment in the ISM of galaxies becomes quickly enhanced.
    
    \item After exploring an exhaustive list of possible sources of systematic error in the measurement of the MZR, we conclude that the presence of the break at $\sim 10^{9.5}$~M$_{\odot}$ cannot be attributed to artifacts arising from any of the following: biases introduced by sample selection, spectroscopic fiber aperture effects, a changing N/O abundance pattern in low mass galaxies, a secondary dependence of metallicity with the SFR, the effects of luminosity weighted HII region ensemble averaging, DIG contamination, nor the statistical methodology used to characterize the metallicty PDF of galaxies and build the MZR.
    
    \item The reason for the discrepancy between our results and other studies in the literature that find a single power-law behaviour in the MZR across the low and intermediate mass regimes lies in systematic differences in the abundances derived using different diagnostics. Both $R23$ SEL diagnostics and the direct method produce a steeper MZR in the low mass regime that lacks the observed break, while the $N2$ and $O3N2$ SEL methods yield an MZR morphology that is consistent with the one we observe with {\tt IZI}, although the latter two methods might not be too reliable in the low abundance regime as discusssed in Section \ref{sec:discussion-test-abundances}.
    
    \item Across the metallicity range that brackets the characteristic mass scale in the MZR, {\tt IZI} abundances of bright local HII regions calculated using the \cite{levesque10} photo-ionization models agree with oxygen abundances from metal RL lines to within $\lesssim 0.1$~dex. These two largely independent methods show practically identical metallicity dependent systematic offsets with respect to the direct method, which can explain the differences in MZR morphology we observe in this work. While the comparison of direct method and RL abundances to stellar abundances of young stars in the same systems could shed light onto which of these diagnostics is less biased, current results in the literature remains inconclusive. Overall, the agreement between {\tt IZI} and RL abundances is encouraging and lends confidence to our result. 
    
    \item High resolution cosmological and zoom-in hydrodynamical simulations of galaxy formation and evolution can reproduce our result. A change of slope in the MZR in the $10^9-10^{10}$~M$_{\odot}$ range that is qualitatively similar to the observed feature reported in this work can be seen in the highest resolution run of the EAGLE simulation \cite{schaye15}, as well as in the zoom-in simulations of \cite{ma16} and \cite{christensen16, christensen18}.
    
    \item In light of theoretical and observational results in the literature we propose that the observed characteristic mass scale in the MZR can be associated to a stellar mass threshold above which galaxies lose the ability to remove metals from their halos. This threshold could be related to a transition in the structure and stability of the star forming ISM, and an associated impact on the efficiency with which feedback can deposit momentum on low column density, super-Eddington, parcels of gas that can be accelerated into galactic outflows.
    
\end{itemize}

The observed shape of the MZR presented in this work has important implications for models of the baryon cycle in galaxies. It will be important to further test this new result observationally, and study its implications for galaxy evolution theory in more detail. Ongoing and future observational campaigns aimed at characterizing the chemistry, structure, and dynamics of the ISM will shed light on the systematics currently plaguing nebular abundance diagnostics. The PHANGS project \citep[Physics at High Angular resolution in Nearby GalaxieS,][]{kreckel18, sun18} is currently mapping the molecular and ionized ISM of nearby ($D<17$~Mpc) spiral galaxies at $50-100$~pc resolution with spatially resolved spectroscopy from ALMA and MUSE, and the stellar content of these systems with {\it HST}. In the future the Local Volume Mapper (LVM) project, part of the fifth realization of the Sloan Digital Sky Survey \citep[SDSS-V,][]{kollmeier17} will use IFU spectroscopy to map the ionized ISM of the MW, the Magellanic Clouds, M31, M33, and a few dozen nearby ($D\lesssim5$~Mpc) star forming dwarf galaxies at spatial resolutions ranging from $0.1$~pc to $100$~pc. This will allow a detailed characterization of the internal ionization structure of HII regions and promises to bring enormous advances in terms of their modeling and the measurement of nebular chemical abundances. These projects will also allow significant progress in our understanding of the star formation process on galactic scales, the injection of metals and feedback into the ISM, and the mixing and redistribution of metals within and outside galaxies.

\vspace{1 cm}

This paper is dedicated to the memory of Professor Michael A. Dopita. Mike was always an enthusiastic, open minded, and friendly researcher, who was open to new ideas and was available and happy to discuss them with younger people. We also thank Jabran Zahid for kindly providing tables of emission line fluxes for the DEEP2 low mass galaxy sample. G.B. was supported by CONICYT/FONDECYT, Programa de Iniciacion, Folio 11150220. A.K. was supported by the CONICYT-FONDECYT fellowship (project number: 3160049). This research made use of NASA's Astrophysics Data System, the NASA/IPAC Extragalactic Database (NED) which is operated by the Jet Propulsion Laboratory, California Institute of Technology, under contract with the National Aeronautics and Space Administration. This work has made use of SDSS DR7 public data. We would like to thank Jarle Brinchman and the team behind the MPA/JHU value added SDSS catalogs. Funding for the SDSS and SDSS-II has been provided by the Alfred P. Sloan Foundation, the Participating Institutions, the National Science Foundation, the U.S. Department of Energy, the National Aeronautics and Space Administration, the Japanese Monbukagakusho, the Max Planck Society, and the Higher Education Funding Council for England. The SDSS Web Site is http://www.sdss.org/. The SDSS is managed by the Astrophysical Research Consortium for the Participating Institutions. The Participating Institutions are the American Museum of Natural History, Astrophysical Institute Potsdam, University of Basel, University of Cambridge, Case Western Reserve University, University of Chicago, Drexel University, Fermilab, the Institute for Advanced Study, the Japan Participation Group, Johns Hopkins University, the Joint Institute for Nuclear Astrophysics, the Kavli Institute for Particle Astrophysics and Cosmology, the Korean Scientist Group, the Chinese Academy of Sciences (LAMOST), Los Alamos National Laboratory, the Max-Planck-Institute for Astronomy (MPIA), the Max-Planck-Institute for Astrophysics (MPA), New Mexico State University, Ohio State University, University of Pittsburgh, University of Portsmouth, Princeton University, the United States Naval Observatory, and the University of Washington.

\clearpage

\begin{deluxetable}{ccccc}

\tabletypesize{\scriptsize}

\tablecaption{Local Volume Dwarf Galaxy Sample MZR \label{tbl-1}{}}

\tablewidth{0pt}

\tablehead{
\colhead{Galaxy} & 
\colhead{$\log{(M_*/M_{\odot})}$\footnotemark} &
\colhead{12+$\log{\rm{O/H}}$} &
\colhead{$\Delta^- \log{\rm{O/H}}$\footnotemark} & 
\colhead{$\Delta^+ \log{\rm{O/H}}$\footnotemark[2]} 
}

\startdata
DDO154 & 6.68 & 7.96 & 0.13 & 0.29 \\
GR8 & 6.62 & 8.12 & 0.18 & 0.35 \\
IC1613 & 6.82 & 8.18 & 0.11 & 0.15 \\
IC2574 & 8.39 & 8.14 & 0.13 & 0.19 \\
IC5152 & 8.11 & 8.34 & 0.13 & 0.13 \\
LeoA & 5.89 & 7.73 & 0.09 & 0.13 \\
M81dwB & 7.19 & 8.24 & 0.13 & 0.20 \\
NGC1569 & 8.07 & 8.27 & 0.14 & 0.13 \\
NGC1705 & 8.13 & 8.29 & 0.12 & 0.11 \\
NGC3109 & 7.41 & 8.36 & 0.38 & 0.11 \\
NGC4214 & 8.55 & 8.29 & 0.07 & 0.12 \\
NGC4449 & 9.29 & 8.31 & 0.03 & 0.10 \\
NGC6822 & 8.10 & 8.18 & 0.11 & 0.16 \\
PegDig & 6.98 & 8.21 & 0.07 & 0.13 \\
SextansA & 6.24 & 7.98 & 0.14 & 0.16 \\
SextansB & 6.86 & 8.16 & 0.10 & 0.08 \\
UGC6456 & 6.90 & 8.11 & 0.22 & 0.36 \\
NGC2363 & 7.91 & 8.20 & 0.11 & 0.13 \\
NGC5408 & 8.19 & 8.31 & 0.13 & 0.16 \\
NGC55 & 8.44 & 8.23 & 0.14 & 0.16 \\
WLM & 6.88 & 8.09 & 0.12 & 0.16
\enddata

\tablenotetext{a}{Taken from Table 1 in \cite{lee06}}
\tablenotetext{b}{Lower and upper bounds of 16\%-84\% confidence interval.}

\end{deluxetable}

\begin{deluxetable}{cccc}

\tabletypesize{\scriptsize}

\tablecaption{SDSS High Mass Sample MZR \label{tbl-2}{}}

\tablewidth{0pt}

\tablehead{
\colhead{$\log{(M_*/M_{\odot})}$} &
\colhead{12+$\log{\rm{O/H}}$} &
\colhead{$\Delta^- \log{\rm{O/H}}$} & 
\colhead{$\Delta^+ \log{\rm{O/H}}$} 
}

\startdata
8.70 & 8.37 & 0.07 & 0.06 \\
8.75 & 8.37 & 0.06 & 0.07 \\
8.80 & 8.37 & 0.06 & 0.07 \\
8.85 & 8.38 & 0.07 & 0.06 \\
8.90 & 8.37 & 0.07 & 0.07 \\
8.95 & 8.40 & 0.06 & 0.07 \\
9.00 & 8.40 & 0.06 & 0.07 \\
9.05 & 8.40 & 0.07 & 0.07 \\
9.10 & 8.41 & 0.07 & 0.06 \\
9.15 & 8.41 & 0.07 & 0.06 \\
9.20 & 8.42 & 0.08 & 0.08 \\
9.25 & 8.42 & 0.08 & 0.08 \\
9.30 & 8.45 & 0.08 & 0.09 \\
9.35 & 8.45 & 0.08 & 0.08 \\
9.40 & 8.46 & 0.09 & 0.10 \\
9.45 & 8.47 & 0.09 & 0.10 \\
9.50 & 8.48 & 0.11 & 0.12 \\
9.55 & 8.48 & 0.11 & 0.12 \\
9.60 & 8.51 & 0.11 & 0.12 \\
9.65 & 8.53 & 0.12 & 0.14 \\
9.70 & 8.53 & 0.13 & 0.13 \\
9.75 & 8.56 & 0.13 & 0.14 \\
9.80 & 8.58 & 0.15 & 0.15 \\
9.85 & 8.62 & 0.15 & 0.15 \\
9.90 & 8.65 & 0.15 & 0.14 \\
9.95 & 8.69 & 0.15 & 0.14 \\
10.00 & 8.71 & 0.14 & 0.12 \\
10.05 & 8.74 & 0.14 & 0.12 \\
10.10 & 8.74 & 0.14 & 0.12 \\
10.15 & 8.78 & 0.12 & 0.11 \\
10.20 & 8.78 & 0.12 & 0.11 \\
10.25 & 8.80 & 0.10 & 0.09 \\
10.30 & 8.80 & 0.10 & 0.09 \\
10.35 & 8.80 & 0.10 & 0.09 \\
10.40 & 8.81 & 0.11 & 0.09 \\
10.45 & 8.82 & 0.09 & 0.08 \\
10.50 & 8.82 & 0.09 & 0.08 \\
10.55 & 8.82 & 0.09 & 0.07 \\
10.60 & 8.82 & 0.09 & 0.07 \\
10.65 & 8.82 & 0.09 & 0.07 \\
10.70 & 8.85 & 0.08 & 0.08 \\
10.75 & 8.84 & 0.07 & 0.06 \\
10.80 & 8.84 & 0.07 & 0.06 \\
10.85 & 8.84 & 0.07 & 0.06 \\
10.90 & 8.86 & 0.06 & 0.07 \\
10.95 & 8.84 & 0.07 & 0.06 \\
11.00 & 8.83 & 0.07 & 0.06 \\
11.05 & 8.84 & 0.07 & 0.06 \\
11.10 & 8.83 & 0.07 & 0.06
\enddata
\end{deluxetable}

\begin{deluxetable}{cccc}

\tabletypesize{\scriptsize}

\tablecaption{SDSS Low Mass Sample MZR \label{tbl-3}{}}

\tablewidth{0pt}

\tablehead{
\colhead{$\log{(M_*/M_{\odot})}$} &
\colhead{12+$\log{\rm{O/H}}$} &
\colhead{$\Delta^- \log{\rm{O/H}}$} & 
\colhead{$\Delta^+ \log{\rm{O/H}}$} 
}

\startdata
7.50 & 8.26 & 0.08 & 0.08 \\
7.75 & 8.28 & 0.07 & 0.06 \\
8.00 & 8.24 & 0.07 & 0.06 \\
8.25 & 8.28 & 0.07 & 0.06 \\
8.50 & 8.28 & 0.07 & 0.06 \\
8.75 & 8.31 & 0.06 & 0.07 \\
9.00 & 8.34 & 0.06 & 0.07 \\
9.25 & 8.34 & 0.07 & 0.06 \\
9.50 & 8.37 & 0.06 & 0.07 \\
9.75 & 8.40 & 0.06 & 0.07 \\
10.00 & 8.41 & 0.07 & 0.07
\enddata

\end{deluxetable}

\begin{deluxetable}{cccc}

\tabletypesize{\scriptsize}

\tablecaption{DEEP2 Low Mass Sample MZR \label{tbl-4}{}}

\tablewidth{0pt}

\tablehead{
\colhead{$\log{(M_*/M_{\odot})}$} &
\colhead{12+$\log{\rm{O/H}}$} &
\colhead{$\Delta^- \log{\rm{O/H}}$} & 
\colhead{$\Delta^+ \log{\rm{O/H}}$} 
}

\startdata
7.25 & 8.18 & 0.07 & 0.10 \\
7.50 & 8.33 & 0.22 & 0.21 \\
7.75 & 8.23 & 0.12 & 0.15 \\
8.00 & 8.28 & 0.16 & 0.20 \\
8.25 & 8.32 & 0.20 & 0.22 \\
8.50 & 8.29 & 0.18 & 0.18 \\
8.75 & 8.36 & 0.19 & 0.21 \\
9.00 & 8.45 & 0.20 & 0.22 \\
9.25 & 8.47 & 0.19 & 0.20 \\
9.50 & 8.57 & 0.19 & 0.20 \\
9.75 & 8.66 & 0.16 & 0.17 \\
10.00 & 8.66 & 0.16 & 0.17
\enddata

\end{deluxetable}

\clearpage

\end{document}